\documentclass{aa}
\usepackage[varg]{txfonts}
\usepackage{natbib,twoopt}
\usepackage[breaklinks=true]{hyperref} 
\bibpunct{(}{)}{;}{a}{}{,} 
\usepackage{graphicx}
\usepackage{subcaption} 
\usepackage{amsmath}    
\usepackage{float}
\usepackage{color}
\usepackage{verbatim}
\usepackage[normalem]{ulem}

\begin{document}
        
        \title{The SRG/eROSITA all-sky survey}
        \subtitle{Identifying the coronal content with HamStar}
        \author{S. Freund\inst{1}\inst{2}
                \and S. Czesla\inst{2}\inst{3}
                \and P. Predehl\inst{1}
                \and J. Robrade\inst{2}
                \and M. Salvato\inst{1}
                \and P. C. Schneider\inst{2}
                \and H. Starck\inst{1}
                \and J. Wolf\inst{1}\inst{4}\inst{5}
                \and\\ J. H. M. M. Schmitt\inst{2}}
        \institute{Max-Planck-Institut f\"ur extraterrestrische Physik, Gie\ss enbachstra\ss e 1, 85748 Garching, Germany
                \email{sfreund@mpe.mpg.de}
                \and
                Hamburger Sternwarte, Universit\"at Hamburg, 21029 Hamburg, Germany
                \and
                Th\"uringer Landessternwarte Tautenburg, Sternwarte 5, D-07778 Tautenburg, Germany
                \and
                Excellenzcluster ORIGINS, Boltzmannstr. 2, 85748 Garching, Germany
                \and
                Max-Planck Institut f\"ur Astronomie, K\"onigstuhl 17, 69177 Heidelberg, Germany}
        \abstract{The first \textit{eROSITA} all-sky survey (eRASS1) performed on board the Spectrum-Roentgen-Gamma mission (SRG) provides more than 900\,000 X-ray sources in the 0.2~--~2.3~keV band located in the western hemisphere.}
        {We present identifications of the eRASS1 sources obtained using our HamStar method, which was designed for the identification of coronal X-ray sources.}
        {HamStar is a Bayesian framework that estimates coronal probabilities for each eRASS1 source based on a cross-match with optical counterparts from \textit{Gaia} DR3. It considers geometric properties, such as angular separation and positional uncertainty, as well the additional properties of fractional X-ray flux, color, and distance.}
        {We identify 138\,800 coronal eRASS1 sources and estimate a completeness and reliability of about 91.5~\% for this sample, which we confirmed with \textit{Chandra} detections. This is the largest available sample of coronal X-ray emitters and we find nearly five times as many coronal sources as in the ROSAT all-sky survey. The coronal eRASS1 sources are made up of all spectral types and the onset of convection and the saturation limit are clearly visible. As opposed to previous samples, rare source types are also well populated. About 10~\% of the coronal eRASS1 sources have a correlated secondary counterpart, which is a wide binary companion or belongs to the same stellar cluster. We also identify 6\,700 known unresolved binaries, and an excess of fast binary periods below 10~d. Furthermore, the binary sequence is clearly visible in a color--magnitude diagram. When combining the coronal eRASS1 sources with rotation modulations from \textit{Gaia} DR3, we find 3\,700 X-ray sources with known rotation periods, which is the largest sample of this kind. We fitted the rotation--activity relation and convection turnover times for our flux-limited sample. 
        We do not detect the low-amplitude fast rotators discovered in the \textit{Gaia} DR3 sample in X-rays.}
        {}
        \maketitle
        
        \section{Introduction}
        Late-type stars with outer convection zones are known to emit X-rays produced in a magnetically heated corona. However,  even
for stars of the same spectral type, the X-ray luminosity can vary by orders of magnitude \citep[see e.g.,][]{vaiana81,schmitt04,gud04,testa15} depending on the ages and rotation periods of the underlying stars \citep{wilson63,skumanich72,preibisch05}. For very young and fast-rotating stars, the X-ray flux does not further increase but saturates outside obvious flares at a level of about $\log(F_X/F_\mathrm{bol})=-3$ \citep{vil84,wright11}. While the rotation period and fractional X-ray flux decrease with time for most stars, close binaries preserve their high activity due to tidal interaction. Therefore, the highest X-ray luminosities are observed for active binaries containing a giant or subgiant component, as in RS CVn systems \citep{wal78,dempsey93}. Also, OB-type stars exhibit high X-ray luminosities, but they are produced by stellar winds and not in a corona \citep{pal81,ber97}.  
        
        Since the {Einstein}
era, X-ray surveys have provided a very useful tool with which to obtain and study unbiased samples of coronal X-ray sources \citep{gio90,sto91}. Examples of important X-ray surveys for stellar science are the \textit{ROSAT} all-sky survey \citep[RASS;][]{vog99,RASS-catalog,freund22} taken in 1990 and the \textit{eROSITA} Final Equatorial-Depth Survey \citep[eFEDS;][]{brunner22,schneider22} that covers an area of 140 square degrees at high Galactic latitudes with high sensitivity. 
        
        
        The main goal of the \textit{eROSITA} X-ray telescope \citep{predehl21} launched in 2019 on board the Spectrum-Roentgen-Gamma mission \citep[SRG;][]{sunyaev21} is to perform the deepest all-sky survey at soft X-ray wavelengths. The release of the first \textit{eROSITA} all-sky survey (eRASS1) executed between December 2019 and June 2020 is presented by \citet{merloni24}. Already, eRASS1 is about 2.5 times more sensitive than RASS and offers unprecedented potential for stellar X-ray astronomy. To harvest this potential, the coronal eRASS1 sources need to be identified and characterized. 
        Characterization of X-ray sources is typically carried out by augmenting the X-ray data with further information from multiwavelength counterparts. All of the geometric aspects of finding the correct counterpart were recently discussed in great detail by \citet{czesla23}, including the probability that the correct counterpart of a given X-ray source is not the nearest neighbor.   It has been found that more reliable identifications can be obtained if further properties of the counterparts, such as the optical magnitude, are included in a Bayesian algorithm, as described for example by \citet{budavari08}, or in a machine learning (ML) approach. Identifications and classifications for large samples of X-ray sources including, but not necessarily restricted to, coronal sources have been presented, for example, for the RASS \citep{salvato18,freund22}, the \textit{XMM-Newton} slew survey \citep{freund18,salvato18} and serendipitous source catalog \citep{tranin22}, the \textit{Chandra} source catalog \citep{yang22,kumaran23}, and eFEDS \citep{salvato22,schneider22}.
        
        For the identification and characterization of the coronal eRASS1 sources, an optical catalog of counterparts with sufficient depth is necessary. To determine the stellar properties of the counterparts, such as spectral type, bolometric flux, and X-ray luminosity, the counterpart catalog should provide optical brightness, color, and parallax. Such a catalog is provided by the \textit{Gaia} mission launched in 2013 \citep{GaiaMission}, and the data taken by \textit{Gaia} are regularly published as intermediate data releases. The most recent release, \textit{Gaia} DR3, is based on the data from the first 34 months of the mission \cite{GaiaEDR3,GaiaDR3}. DR3 contains positions and magnitudes for 1.8 billion sources, and also parallaxes, proper motions, and colors are provided for 1.5 billion sources brighter than 21$^\mathrm{}$ mag in \textit{Gaia}'s G-band. Therefore, \textit{Gaia} provides for the first time a catalog that contains the optical counterparts to all coronal X-ray emitters detected by \textit{eROSITA} and delivers parallax measurements for all these sources to the extent that the \textit{Gaia} catalog is complete. \textit{Gaia} DR3 is thought to be highly complete in the range $7<G<19$~mag, but there are known incompleteness issues for the brightest and faintest sources, as well as in crowded regions. 
        
        Here, we describe a new identification procedure called HamStar, which is optimized for the identification of coronal X-ray emitters, and present the coronal content of eRASS1 obtained by this procedure. We decided to use a Bayesian framework instead of a ML approach because it delivers results of  similar quality and provides well-calibrated coronal probabilities for each X-ray source \citep{schneider22}. Although optimized for coronal sources, HamStar also identifies some early-type sources that have similar properties. We do not aim to identify Galactic X-ray sources produced by accreting processes, such as cataclysmic variables (CVs; see Schwope et al. in prep. for an identification of these source types).
        Our paper is structured as follows; in Sect.~\ref{sec: Catalogs}, we outline the X-ray and optical catalog that we used. We then describe the matching procedure with HamStar in Sect.~\ref{sec: identification problem}. We present our results in Sect.~\ref{sec: Results} and discuss the properties of the coronal eRASS1 sources in Sect.~\ref{sec: Properties of the stellar eRAS1 sources}. In Sect~\ref{sec: Error discussion and caveats}, we specify the limitations of our identifications. Finally, we draw our conclusions and provide an outlook in Sect.~\ref{sec: Conclusions}.

        \section{Data sources}
        \label{sec: Catalogs}
        
        \subsection{First eROSITA all-sky survey}
        The eRASS1 catalog contains sources located in the half sky west of the Galactic center and is divided in a main and a hard catalog of sources detected in the 0.2~--~2.3~keV and 2.3~--~5~keV bands, respectively, as well as a supplementary catalog that contains sources with a low detection likelihood. As coronal sources are known to be soft X-ray emitters, we adopted the approximately 900\,000 point-like sources of the main catalog as the basis 
        for the identification of coronal X-ray sources. 
        A detailed description of the eRASS1 catalog is provided by \citet{merloni24} and we summarize the most important aspects in the context of this work. 
        
\begin{figure}[t]
                \includegraphics[width=\hsize]{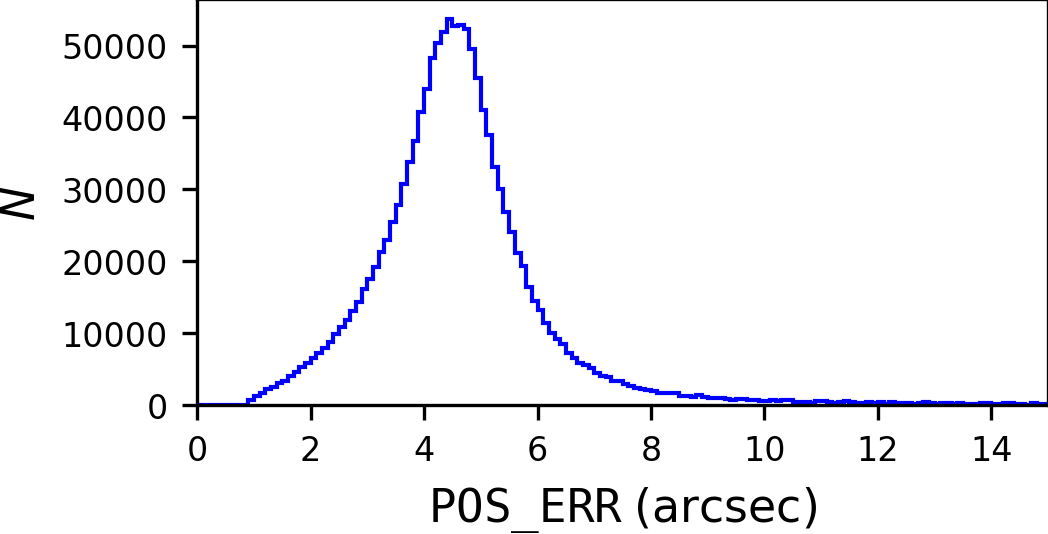}
                \caption{Distribution of the ``$1\sigma$'' positional uncertainty for the eRASS1 sources}
                \label{fig: eRASS1 POS_ERR}
        \end{figure}

        Due to the SRG scanning law, sources near the ecliptic equator are observed six times 
        during one day, while sources at higher ecliptic latitudes are observed for longer periods
        of time. As a consequence, the vignetting-corrected exposure time in eRASS1 varies 
        between typically 150~s near the ecliptic equator and several thousand seconds 
        near the ecliptic poles.  To obtain energy fluxes from the vignetting-corrected 
        count rates, we need to apply an energy conversion factor (ECF).  The ECF for coronal 
        sources depends somewhat on the assumed spectral model. We estimated ECFs adopting APEC models with different coronal temperatures and find that an ECF of $8.5\times 10^{-13}$~erg\,cm$^{-2}$\,cnt$^{-1}$ is appropriate for most coronal emitters detectable in eRASS1. 
        This choice of ECF results in a sensitivity that varies for most eRASS1 sources between $2.5\times10^{-14}$ and $8\times10^{-14}$~erg\,s$^{-1}$\,cm$^{-2}$ with a median of about $5.5\times10^{-14}$~erg\,s$^{-1}$\,cm$^{-2}$. The ``$1\sigma$'' positional uncertainty (\texttt{POS\_ERR}) of the eRASS1 sources depends, primarily, on the number of detected counts and is on average about 4.6~arcsec, which is roughly the pixel size of the eROSITA CCDs. Figure~\ref{fig: eRASS1 POS_ERR}  shows the distribution of the positional errors for the eRASS1 sources.
        
        The main catalog contains point sources with a detection likelihood of $>6$ and about 14\% of these sources are expected to be spurious detections \citep{seppi22}. The probability of a source being spurious depends on the detection likelihood, and furthermore, the eRASS1 catalog contains flags for likely spurious sources, such as very bright X-ray point sources or supernova remnants in the vicinity. Except for sources near a star cluster, where we expect to find many coronal X-ray emitters, the number of reasonable coronal counterparts is reduced for the flagged sources(see Sect.~\ref{sec: Distribution of the stellar fraction}), and we refer to sources with a flag other than the stellar cluster flag\footnote{namely \texttt{FLAG\_SP\_SNR}, \texttt{FLAG\_SP\_BPS}, \texttt{FLAG\_SP\_LGA}, and \texttt{FLAG\_GC\_CONS}} as likely spurious detections. Optically very bright sources may produce spurious eRASS1 detections because of the accumulation of optical protons over the integration time of a CCD pixel, which is known as optical loading. Sources that might be affected are flagged in the eRASS1 catalog (\texttt{FLAG\_OPT}).
        
        At the time of writing, preliminary data combined from the first four all-sky surveys are already available within the \textit{eROSITA} consortium. Similar to  the approach described in \citet{merloni24} for eRASS1, a source catalog is constructed from these data. In the following, we use the eRASS:4 source catalog version 221031, which is processed by the eSASS pipeline version c020 to improve our prior knowledge of the properties of coronal sources detected in eRASS1 (see Sect.~\ref{sec: Bayes map construction}). 
        
        \subsection{Gaia DR3}
        The astrometric data processing of \textit{Gaia} DR3 and the ensuing results are discussed in detail by \citet{Lindegren21}. In summary, a color-dependent calibration of the point-spread function is applied to nearly 585 million sources with a well-determined color in \textit{Gaia} DR2. This five-parameter solution cannot be adopted for about 882 million sources with insufficient color quality, which applies in particular to either very bright ($G < 4$~mag) or relatively faint ($G \gtrsim 16$~mag) sources. Instead, a pseudo-color is estimated as the sixth astrometric parameter. The accuracy of the positions and parallaxes typically varies between 0.02 and 0.03~mas for sources with $G < 15$~mag and between 0.4 and 0.5~mas at $G=20$~mag, respectively. Sources with a five-parameter solution generally have higher accuracies. 
        
        \textit{Gaia} DR3 provides mean G-band fluxes for nearly all sources in the \textit{Gaia} catalog and mean BP and RP fluxes for about 85~\% of them, while most of the sources with missing colors are faint. At $G=17$~mag, an accuracy of 1, 12, and 6~mmag in the G, BP, and RP bands is 
        obtained, respectively; however, the uncertainties strongly decrease and increase for brighter and fainter sources, respectively; more details on these issues can be found in \citet{Riello21}.
        
        Furthermore, \textit{Gaia} DR3 provides low-resolution spectra for nearly 320 million and high-resolution RVS spectra for about 1 million sources. Astrophysical parameters and radial velocities are estimated for more than 475 million and 33 million sources, respectively, and samples of variable sources and non-single stars contain 10.5 million and 0.8 million objects, respectively. Based on the magnitude measurements of the various focal plane transits during the \textit{Gaia} mission, \citet{distefano23} present rotation modulations for 474\,000 \textit{Gaia} DR3 late-type main sequence stars.

        \section{The HamStar identification procedure}
        \label{sec: identification problem}
        The HamStar algorithm is largely based on the method described by \citet{schneider22} and \citet{freund22}, and we outline the basic aspects and improvements in the following.
        Specifically, HamStar searches for so-called eligible coronal counterparts within a search radius of five times the positional uncertainty of the eRASS1 source, and we define and explain our motivation for the selection criteria of our eligible coronal counterparts in Sect.~\ref{sec: data selection}. Throughout this paper, we take any counterpart as a candidate association and use the terms identification and association synonymously. We describe our estimation of the coronal probabilities in Sect.~\ref{sec: stellar probabilities}, and discuss our construction of the Bayes map in Sect.~\ref{sec: Bayes map construction}. 
        
        \subsection{Data selection}
        \label{sec: data selection}
        For every given eRASS1 X-ray catalog source, we select a sample of eligible coronal counterparts by restricting
        the cross-match to \textit{Gaia} DR3 sources that are brighter than $G=19$~mag and have a significant parallax (i.e., $\frac{\pi}{\sigma_\pi}>3$). 
        Furthermore, we require the sources to have flux measurements in both the BP and RP bands, because the combined color and parallax 
        information enables us to place the sources in a color--magnitude diagram (CMD; see Sect.~\ref{sec: color-magnitude diagram}).   As HamStar
        focuses on the identification of  ``normal'' (i.e., non-degenerate) stars, the possible locations of such objects in a  \textit{Gaia}-based CMD are
        relatively restricted; furthermore, this information is necessary for our weighting scheme using a  Bayes map (see Sect.~\ref{sec: Bayes map construction}). Due to the known incompleteness of \textit{Gaia} DR3 at the bright end, we added some counterparts from the Tycho2 catalog with measurements in the $B_T$ and $V_T$ bands and parallaxes from \textit{Hipparcos,} and estimated the magnitudes in the \textit{Gaia} photometric bands, adopting the conversion factors of the \textit{Gaia} DR3 online documentation \citep{GaiaDocPhot}. 
        
        \begin{figure}[t]
                \includegraphics[width=\hsize]{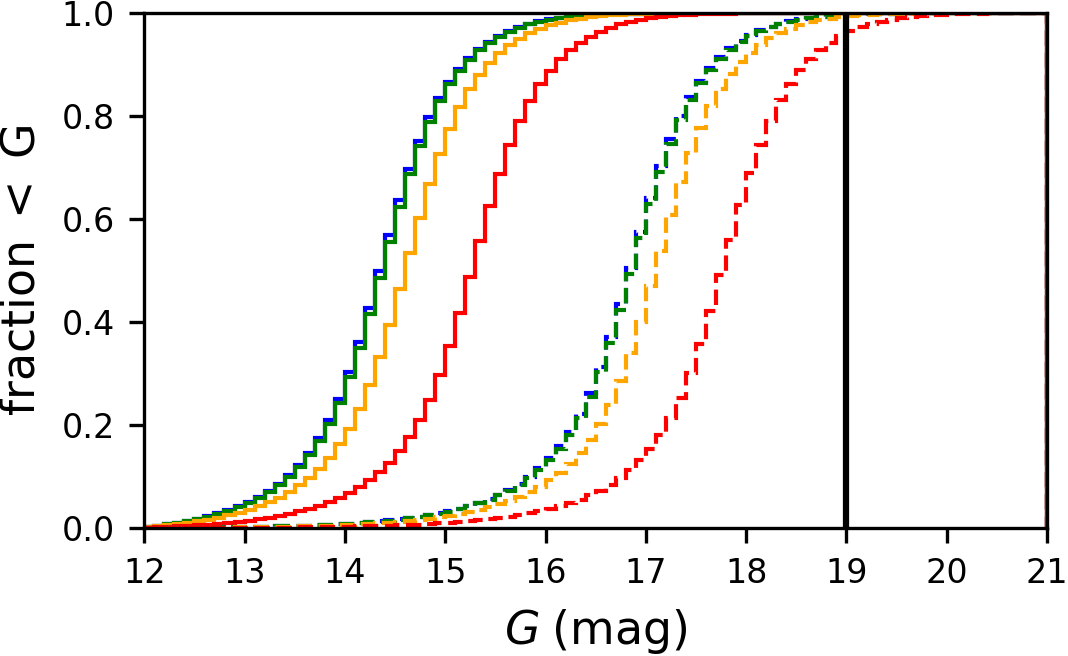}
                \caption{Cumulative distribution of the limiting G-band magnitudes for the eRASS1 sources. For the blue, green, orange, and red lines, counterparts of spectral type F ($BP-RP = 0.6$~mag), G ($BP-RP = 0.85$~mag), K ($BP-RP = 1.4$~mag), and M ($BP-RP = 2.5$~mag) are considered, respectively. Maximal fractional X-ray fluxes of $\log(F_X/F_\mathrm{bol})=-3$ and $\log(F_X/F_\mathrm{bol})=-2$ are assumed for the solid and dashed lines, respectively.}
                \label{fig: limiting magnitudes}
        \end{figure}
        
        We looked for possible incompleteness introduced by these selection criteria, as follows: the limiting brightness of reasonable coronal eRASS1 counterparts in the G-band depends on the X-ray flux of the eRASS1 source considered as well as on the X-ray over bolometric\footnote{For the bolometric corrections and color conversions throughout this paper, we adopted a table that is based on \citet{pecaut13} and regularly updated under \url{http://www.pas.rochester.edu/~emamajek/EEM_dwarf_UBVIJHK_colors_Teff.txt} (current version 2022.04.16).} flux ratio limit and the color of the counterpart. We estimated the expected G-band magnitudes for every eRASS1 source, considering different colors and fractional X-ray fluxes, and show the resulting cumulative distributions in Fig.~\ref{fig: limiting magnitudes}. The quiescent X-ray flux of coronal sources is known to saturate at about $\log(F_X/F_\mathrm{bol}) = -3$ \citep{vil84,wright11}, and therefore coronal counterparts of eRASS1 sources of all spectral types are generally expected to be significantly brighter than our chosen cutoff at $G=19$~mag. Even M-type stars detected at $\log(F_X/F_\mathrm{bol}) = -2$, as possibly observed due to a large flare or  due to statistical variations, are expected to be brighter than $G=19$~mag for more than 95~\% of the eRASS1 sources. Therefore, we expect to miss only very few true coronal associations due to our brightness cut and only if the source is detected during a strong flare; furthermore, this limitation is only relevant for late to very late M dwarfs.
        
        \begin{figure}[t]
                \includegraphics[width=\hsize]{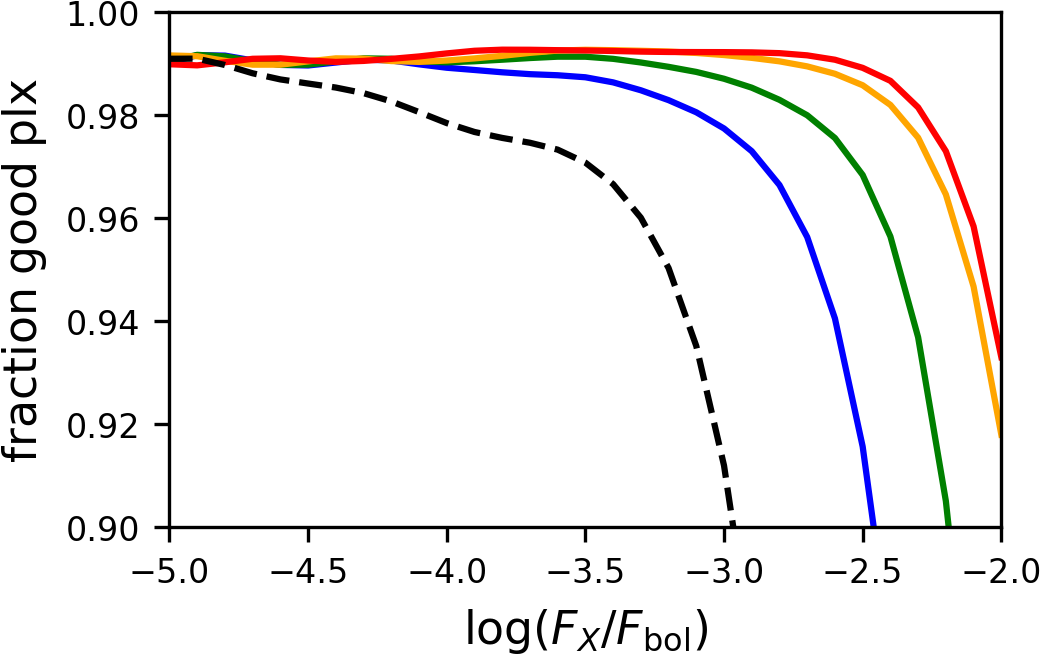}
                \caption{Distribution of coronal eRASS1 counterparts with reliable parallax as a function of the X-ray over bolometric flux. For the blue, green, orange, and red lines, dwarfs of spectral type F ($BP-RP = 0.6$~mag), G ($BP-RP = 0.85$~mag), K ($BP-RP = 1.4$~mag), and M ($BP-RP = 2.5$~mag) are considered as counterparts. Giants located 5~mag above the main sequence and with $BP-RP = 1.2$~mag are shown by the dashed black line.}
                \label{fig: fraction good plx}
        \end{figure}
        We might also miss some correct identifications because of counterparts with a parallax of smaller than three times its error or completely without a parallax measurement in \textit{Gaia} DR3. In addition to the expected G-band magnitudes, we estimated for every eRASS1 source the expected distance of a coronal counterpart, adopting different fractional X-ray fluxes and CMD positions. We then checked the fraction of \textit{Gaia} DR3 sources with a similar brightness that satisfy our parallax cutoff criteria and sum this fraction over all eRASS1 sources. Figure~\ref{fig: fraction good plx} shows the resulting fraction of counterparts with a good parallax. This fraction does not exceed 99\% because of sources without a parallax measurement. For large fractional X-ray fluxes, this fraction decreases because these sources can be detected up to larger distances, and furthermore, the sources are optically fainter, causing larger parallax uncertainties. The parallax quality of F- and G-type dwarfs in particular rapidly decreases for high fractional X-ray fluxes; however, activity levels above the saturation limit are very rare for these sources (cf. Sect.~\ref{sec: X-ray over bolometric flux}). We therefore expect more than 93\% of the counterparts to fulfill our parallax criteria even for the most active M-dwarfs. The fraction of sources with an unreliable parallax might be larger for giants because they can be detected up to larger distances. However, the fractional X-ray flux of giants typically does not reach the saturation limit, except for RS CVn-type systems usually located in the subgiant regime.  For most giants detected in eRASS1, we therefore expect the fraction of sources with a good parallax to be above 95\% and very few missed counterparts due to the chosen parallax cutoff. We further emphasize that a reliable parallax measurement is key to our weighting of the counterparts (see Sect.~\ref{sec: Bayes map construction}) and to our analysis of the properties of coronal X-ray sources.
        
        It is important to keep in mind that the eligible coronal counterparts selected in this fashion may also contain some associations to X-ray sources produced by accreting processes such as CVs or even neutron stars in addition to the coronal sources. These objects cannot be generally identified by their optical 
        properties alone, and they need to be filtered out; for example, by the X-ray over optical flux ratio, which is typically far higher than that of coronal sources (see Sect.~\ref{sec: non-coronal contamination}).

        \subsection{Coronal probabilities}
        \label{sec: stellar probabilities}
        For each of the $N_X$ eRASS1 sources, we have a number of $N_O$ eligible optical coronal counterparts; in principle, we could consider the whole
        \textit{Gaia} catalog, but in practice we consider only those eligible counterparts that are located closer than the  ``$5\sigma$'' positional uncertainty (\texttt{POS\_ERR}) to the eRASS1 source.   Thus,
        57~\% of the eRASS1 sources (predominantly located outside the Galactic plane) have $N_O$ = 0; that is, for these sources there exists no eligible coronal counterpart.
        If $N_O > 0$, we test the hypotheses $\big(H_{ij}\big)_{j=1,N_O}$ that the $i^\mathrm{th}$ eRASS1 source is associated with the $j^\mathrm{th}$ optical counterpart, while the
        hypothesis $H_{i0}$ states that the $i^\mathrm{th}$ eRASS1 source has no association in the catalog of optical counterparts. 
        The association probability is then defined by the following parameters. The most important values for the identification are the proper -motion-corrected angular separation $r_{ij}$ between the $i^\mathrm{th}$ X-ray source and the $j^\mathrm{th}$ counterpart, and the positional error $\sigma_i$ of the eRASS1 source. Given the submilliarcsecond accuracy of the \textit{Gaia} positions, the uncertainties in the positions of the optical counterparts can be neglected.
        In order to assess the probability of pure random associations, we need 
        to consider the spatial counterpart density $\eta$ in the vicinity of the eRASS1 source, for which we estimated the density of the eligible coronal counterparts in HEALPix \citep{healpix05} pixels with a resolution of less than half a degree. 
        
        As we performed the cross-match with only eligible coronal counterparts and the majority of eRASS1 sources are expected to be caused by AGN, the correct identification of most sources is not within our counterpart sample. Furthermore, about 14~\% of the eRASS1 sources are expected to be spurious, which means that they do not have a counterpart in any catalog. Therefore, we applied a prior considering the fraction of real (i.e., not spurious) X-ray detections $p_r$ and the expected fraction of coronal sources within the real detections $p_c$ (see Equations~3 and 4 of \citet{freund22}).
        The fraction of spurious eRASS1 detections is a function of the detection likelihood provided by \citet{seppi22} in their Table~3, while the coronal fraction is a function of the position on the sky (see Sect.~\ref{sec: Distribution of the stellar fraction}). As described by \citet{freund22}, the best value of the coronal fraction can be estimated by maximizing the likelihood of the matching configuration (see their Equation~9), and we divided the eRASS1 half sky in 96 Healpix pixels and optimized the coronal fraction in every pixel. We excluded eRASS1 sources flagged as likely spurious detections from the estimation of the coronal fraction because the number of reasonable coronal counterparts is generally reduced for these sources (see Sect.~\ref{sec: Distribution of the stellar fraction}). 
        
        Following the derivation from \citet{freund22}, the matching probability that the $i^\mathrm{th}$ eRASS1 source is associated with the $j^\mathrm{th}$ optical counterpart is then derived from the following expression: 
        \begin{equation}
                \label{prob_match}      
                p_{ij} = \frac{p_r p_c/(1-p_r p_c) \cdot L_{ij}}{\eta\Omega + p_r p_c/(1-p_r p_c) \cdot \sum_{k=1}^{N_O} L_{ik}},
        \end{equation}
        with the likelihood
        \begin{equation}
                L_{ij} = \frac{2}{\sigma_i^2} \cdot \exp\left(-\frac{r_{ij}^2}{2\sigma_i^2}\right) \cdot B_{ij}\left(\frac{F_X}{F_G}, BP-RP, d\right),
        \end{equation}
        where $\Omega$ is the area of the full sky and $B_{ij}$ is the Bayes factor that considers additional properties; namely, the X-ray to G-band flux ratio $F_X/F_G$, the color $BP-RP$, and the distance $d$ of the counterpart. If no additional properties are considered (i.e., $B_{ij}=1$), we refer to this probability as the geometric matching probability $p_{ij,geo}$. The probability that any one of the eligible counterparts is the correct identification and that the eRASS1 source is produced by coronal emission is estimated as 
        \begin{equation}
                \label{prob_stel}       
                p_\mathrm{coronal} = \sum_{k=1}^{N_O} p_{ik} \frac{p_r p_c/(1-p_r p_c) \cdot \sum_{k=1}^{N_O} L_{ik}}{\eta\Omega + p_r p_c/(1-p_r p_c) \cdot \sum_{k=1}^{N_O} L_{ik}}.
        \end{equation}
        
        As is apparent from Eq.~\ref{prob_match},       for counterparts with separations larger than the positional uncertainty, the probabilities go rapidly to zero, 
        which is why we only consider counterparts within $5\sigma$ of the eRASS1 uncertainty. 
         
        \subsection{Bayes map construction}
        \label{sec: Bayes map construction}
        Due to the relatively large positional uncertainties of X-ray sources (typically several arcseconds), the search radii of nearly 20\% of the eRASS1 sources contain multiple possible counterparts. Therefore, the geometric information of the match is often not sufficient to identify the correct association, and, instead, additional physical properties need to be considered.  With the help of a so-called 
        Bayes map, we assign a Bayes factor to all of the geometric counterparts.   This Bayes factor is computed using the physical characteristics
        of true coronal X-ray sources and greatly helps to distinguish such true coronal counterparts from 
        chance alignments.   To obtain the properties of only spurious identifications in the region of eRASS1 sources, we randomly shifted all eRASS1 sources between 10 and 20 arcmin; in this fashion, the global properties (i.e., space density of the sample) remain, but all individual
        matches are destroyed. We performed the same matching procedure to both the real and the shifted sources, excluding likely spurious detections, and refer to the associations obtained in this way as the control set in the following.
        
        We deduced the properties of true coronal X-ray emitters from a training set, which we constructed from sources with very good positional matches. Specifically, we selected counterparts with a geometric matching probability of $p_{ij,geo}>0.95$. As there are a few other source types in the sample of eligible coronal counterparts, the geometrically selected sample contains some associations that,  although likely responsible
for the X-ray emission, are clearly not coronal. Therefore, we excluded sources with a large X-ray luminosity ($L_X > 10^{32}$~erg/s), a high X-ray over bolometric flux\footnote{adopting the empirical relation\\ $\log(F_X/F_G) > \begin{cases} (BP-RP)\times 1.7 - 3.58 & :\; BP-RP<0.7~\mathrm{mag}\\ (BP-RP)\times 0.727 - 2.9 & :\; BP-RP>0.7~\mathrm{mag}\\ \end{cases}$}, and sources located more than 1.5~mag below the main sequence. Furthermore, we excluded counterparts brighter than $G=4.5$~mag because these are likely affected by optical loading. We found that sources in the Orion region are somewhat over-represented in our training set, and therefore, we excluded sources within 80~pc from the center of the Orion cloud from our training and control set. 
        
        \begin{figure}[t]
                \includegraphics[width=\hsize]{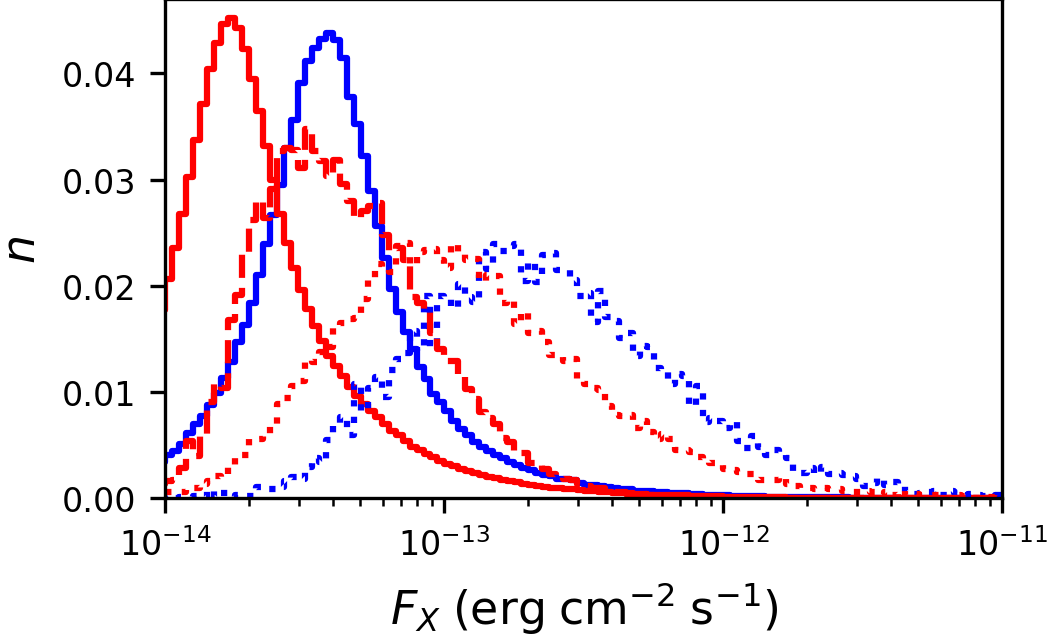}
                \caption{Comparison of the X-ray fluxes in the training set and the complete catalog. The solid blue and red lines show the flux distributions of all eRASS1 and eRASS:4 sources, while X-ray fluxes of the eRASS1 and eRASS:4 training set are shown by the dotted blue and red lines, respectively. For the dashed red histogram, the eRASS:4 training set sources are weighted (see text for details). All distributions are normalized to the number of sources in the sample.}
                \label{fig: Fx distribution training set}
        \end{figure}
        
        In this fashion, we selected 10\,298 training-set sources from eRASS1. In Fig.~\ref{fig: Fx distribution training set}, we compare the brightness distribution of the training set with the complete eRASS1 sources. Figure~\ref{fig: Fx distribution training set}  shows that the eRASS1 training-set sources are generally brighter than the complete eRASS1 sample, because the positional uncertainty of eRASS1 sources is strongly correlated with the number of detected counts, and therefore relatively  X-ray-bright sources will have positional associations that are more secure; even the 35\,606 training set sources from the more sensitive eRASS:4 survey are still brighter than an average eRASS1 source. 
        
\begin{figure}[t]
                \includegraphics[width=\hsize]{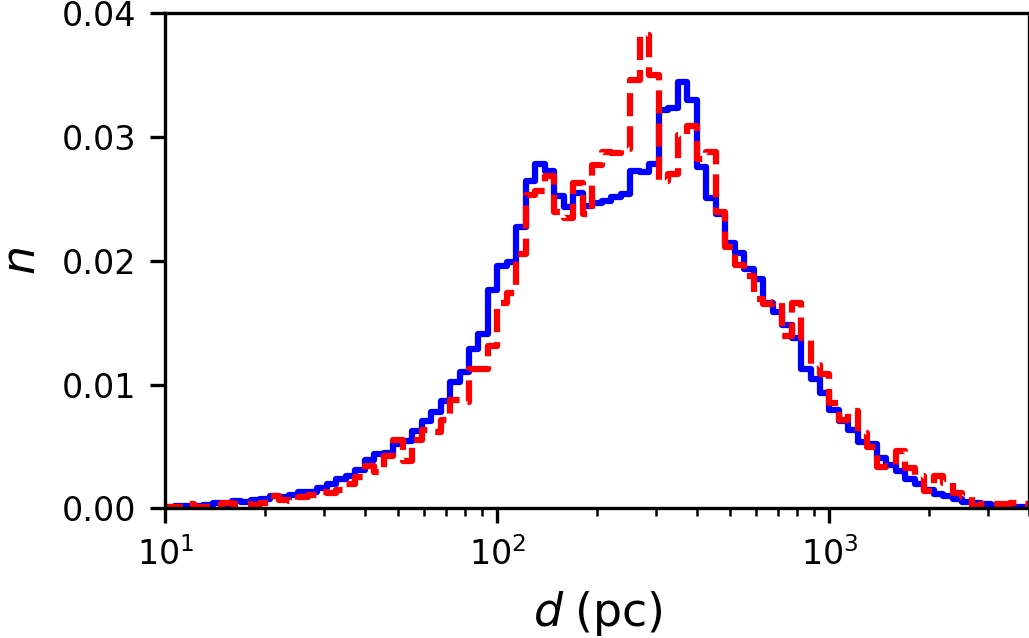}
                \caption{Comparison of the distances for the weighted eRASS:4 training-set sources (red dashed line) and the coronal eRASS1 sources (blue solid line)}
                \label{fig: distance distribution training set}
        \end{figure}

        However, X-ray-bright training-set sources bias the counterpart selection because they also typically have brighter and nearer optical counterparts. 
        We therefore applied different weighting factors to the training-set sources. To this end, we estimated the minimal number of detectable counts from the exposure time for every training-set source, and  from that we derived the maximal distance $d_\mathrm{max}$ up to which the considered source could have been detected. 
        We then adopted for each training-set source a weighting factor estimated as 
        \begin{equation}
                W = \frac{4/3 \pi d^3}{4/3 \pi d_\mathrm{max}^3} = \frac{V}{V_\mathrm{max}},
        \end{equation}
        where $d$ is the distance of the counterpart and $d_\mathrm{max}$ is the maximum distance, and $V$ and $V_\mathrm{max}$ are the corresponding volumes. 
        In this way, a training-set source near the detection limit obtains a larger weight than a bright source. By comparing the solid blue and dashed red lines in Fig.~\ref{fig: Fx distribution training set}, it is obvious that the weighted training set from eRASS:4 recovers the X-ray flux distribution of the eRASS1 catalog much better. Also, the distances of the weighted eRASS:4 training-set sources are similar to those of all eRASS1 sources identified as coronal, as shown in Fig.~\ref{fig: distance distribution training set}. Therefore, we applied the weighted training set from eRASS:4 to describe the properties of true coronal X-ray emitters and we flag the eRASS1 counterparts to these sources in our catalog (see Appendix~\ref{sec: column description}).
        
        The Bayes map was then derived by the ratio between the probability density functions (PDFs) of the properties from the weighted eRASS:4 training set and those of the control set sources. As the properties of the background sources change with the sky position, we divided the control set in the bins shown in Fig.~4 of \citet{freund22} and estimated different Bayes maps in the different areas of the sky. For eRASS1, we constructed three-dimensional Bayes maps that consider the X-ray over G-band flux, $BP-RP$ color, and the distances of the counterparts. Throughout this study, we estimated the distance as the inverse of the parallax. As the coronal eRASS1 sources are rather bright and nearby in the Gaia context,  for most
sources this estimation agrees with that resulting from more sophisticated methods \citep[e.g., by][]{bailer-jones21} within 5\% and our identifications are consistent to more than 99.4~\%. Therefore, we decided to apply the model-independent parallax inversion. In Appendix~\ref{sec: Bayes maps}, we show and discuss examples of the Bayes map.

        \section{Results}
        \label{sec: Results}
        We applied our HamStar procedure described in Sect.~\ref{sec: identification problem} to the point-like sources of the eRASS1 main catalog. In Sect.~\ref{sec: Distribution of the stellar fraction}, we discuss the number of coronal eRASS1 sources estimated by HamStar, and the completeness and reliability of the our sample is discussed in Sect.~\ref{sec: completeness and reliability}. Due to its superior positional accuracy, detections with the \textit{Chandra} X-ray Observatory provide an excellent opportunity to gauge the validity our identifications and we compare our counterparts with \textit{Chandra} detections in Sect.~\ref{sec: Comparison with Chandra detections}. Salvato et al. (in prep.) present identifications of the counterparts (regardless of their extragalactic or Galactic nature) to the eRASS1 point sources using NWAY \citep{salvato18}, which adopts the Legacy Survey DR10 (LS10), CatWise2020, and \textit{Gaia} DR3 as secondary catalogs. As a reminder, NWAY provides different quantities that can be used to assess the reliability of an association; in particular, for each source (eROSITA sources in this case) in a primary catalog,  it provides   the probability p\_any of having a counterpart present in the secondary catalog (in this application either LS10, CatWise2020, or Gaia DR3), and for each source in the secondary catalog within a certain distance from the primary, the probability p\_i of it being the right counterpart. While the area covered with LS10 leads to the most reliable results, the identifications using counterparts from \textit{Gaia} DR3 allow a direct comparison. Furthermore, LS10 does not cover the Galactic plane, which is of particular importance for coronal X-ray sources. Therefore, we compare our results with the identification from Salvato et al.(in prep.) in Sect.~\ref{sec: comparison with identifications from NWAY}. We outline the publication of our matching catalog in Sect.~\ref{sec: publication of the stellar identifications}.  
        
        \subsection{Number of coronal eRASS1 sources and the coronal fraction}
        \label{sec: Distribution of the stellar fraction}
        By summing the coronal probabilities of all eRASS1 sources, we expect a total number of 139\,700 coronal X-ray emitters detected in eRASS1, which corresponds to a coronal fraction of 15.5~\% of all point-like eRASS1 main sources. The coronal fraction of eRASS1 is smaller than in the shallower RASS, in which 24.9~\% of the sources are coronal \citep{freund22}, but larger than in the more sensitive eFEDS survey \citep[7.5~\%;][]{schneider22}  located at high Galactic latitudes. 
        
\begin{figure}[t]
                \includegraphics[width=\hsize]{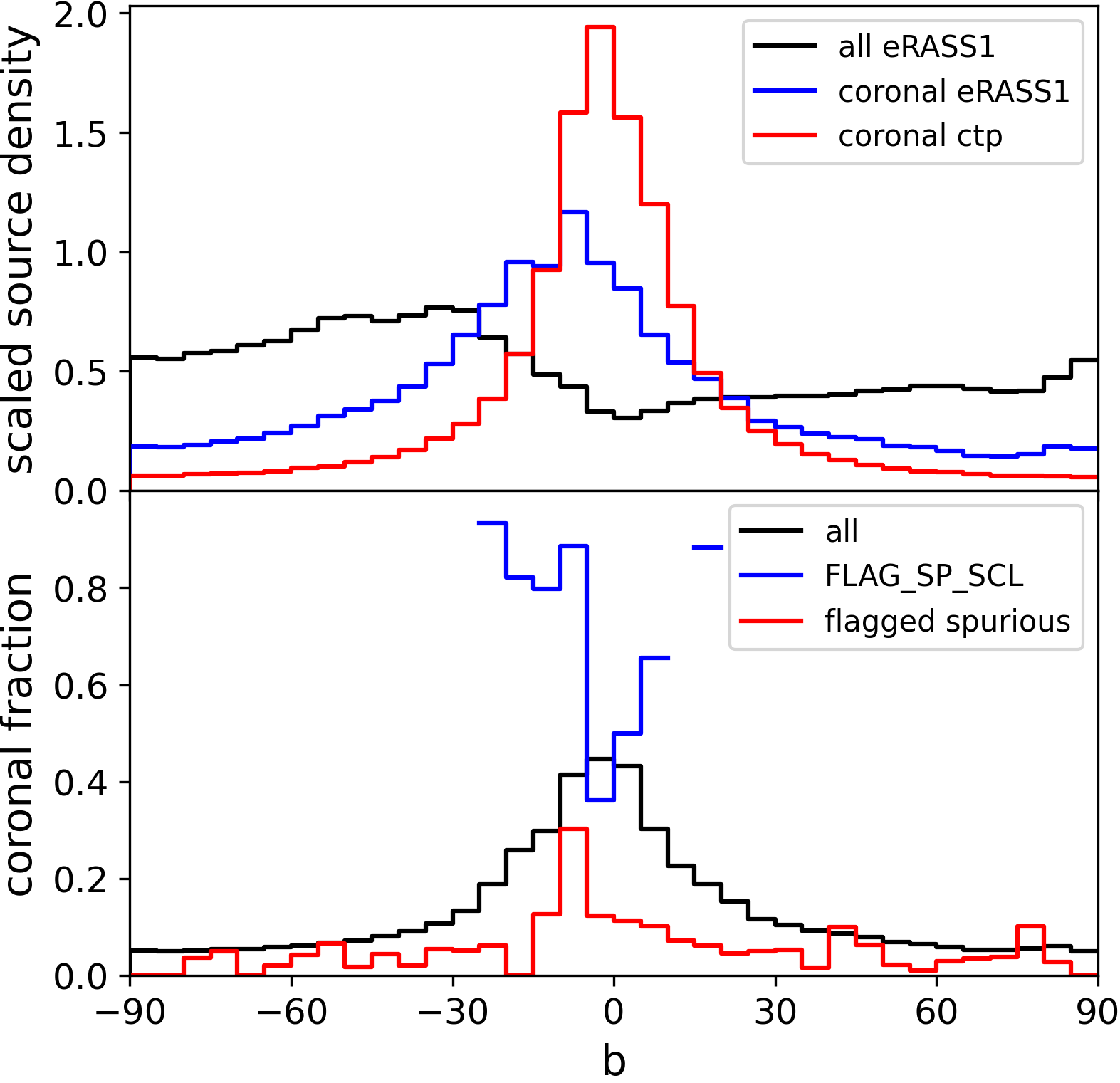}
                \caption{Distribution of the source density and the coronal fraction as a function of Galactic latitude. The scaled source density of all eRASS1 sources, the coronal eRASS1 sources, and the eligible coronal counterparts are shown in the top panel by the black, blue, and red lines, respectively. The source densities are scaled by the area covered by the bins and the total number of objects in the different samples. The bottom panel shows the distribution of the coronal fraction. The fraction of stars for all eRASS1 sources estimated with the Bayes map is indicated by the solid black line. The red and blue lines show the coronal fraction for the sources flagged in the eRASS1 catalog as likely spurious detections and as sources in stellar clusters, respectively.}
                \label{fig: Distribution stellar fraction}
        \end{figure}

        In Fig.~\ref{fig: Distribution stellar fraction} we compare the Galactic latitude distribution of all eRASS1 sources with the latitudes of the our coronal identifications. The overall density of eRASS1 sources decreases towards the Galactic plane, because extragalactic sources are absorbed by the interstellar medium. The largest density of eRASS1 sources is reached near the southern ecliptic pole at $b\approx -29.8^\circ$ because this is the region with the deepest exposure. On the other hand, the density of the coronal eRASS1 sources increases with decreasing Galactic latitude but not as strongly as for the eligible coronal counterparts, because the coronal eRASS1 sources  typically have smaller distances, and are therefore less strongly concentrated on the Galactic plane. Consequently, the coronal fraction is not constant over the sky but increases at low Galactic latitudes, reaching a value of more than 45~\% in the Galactic plane, while only about 5\% of the eRASS1 sources at the Galactic poles are stars. The coronal fraction is substantially smaller for sources that are flagged in the eRASS1 catalog, which confirms that many of these sources are likely spurious detections. The only exception is the spurious flag \texttt{FLAG\_SP\_SCL} of eRASS1 sources located in overdensities associated with Galactic star clusters. Sources in stellar clusters are young, and therefore many of these sources are detected in X-rays resulting in a high coronal fraction of eRASS1 sources with \texttt{FLAG\_SP\_SCL}. 
        
        The number of expected coronal eRASS1 sources increases to 154\,200 if only geometric properties are considered. The difference between the coronal fractions estimated geometrically and those estimated with the Bayes map is largest in the Galactic plane. The deviation is caused by noncoronal sources in our sample of eligible coronal counterparts; for example, some accreting objects might pass our brightness and parallax cutoffs. As these are the true associations of the X-ray emission, they correspond geometrically well with the eRASS1 position, but are excluded by the Bayes map because their properties differ from those expected for coronal X-ray emitters. Similarly, a few extragalactic sources have spuriously high significant parallaxes in \textit{Gaia} DR3 and might be part of our counterpart sample. However, the X-ray to optical flux ratio of CVs and extragalactic objects is generally much higher than that for coronal sources, and therefore the contamination of these source types in our sample of coronal eRASS1 sources is substantially reduced by the Bayes map (cf. Sect.~\ref{sec: non-coronal contamination}). The discrepancy between the geometrically and Bayesian expected number of coronal eRASS1 sources decreases for larger detection likelihoods, which might indicate that we additionally miss some true coronal identifications for faint X-ray sources (cf. Sect.~\ref{sec: Comparison with Chandra detections}). 
        
\begin{figure}[t]
                \includegraphics[width=\hsize]{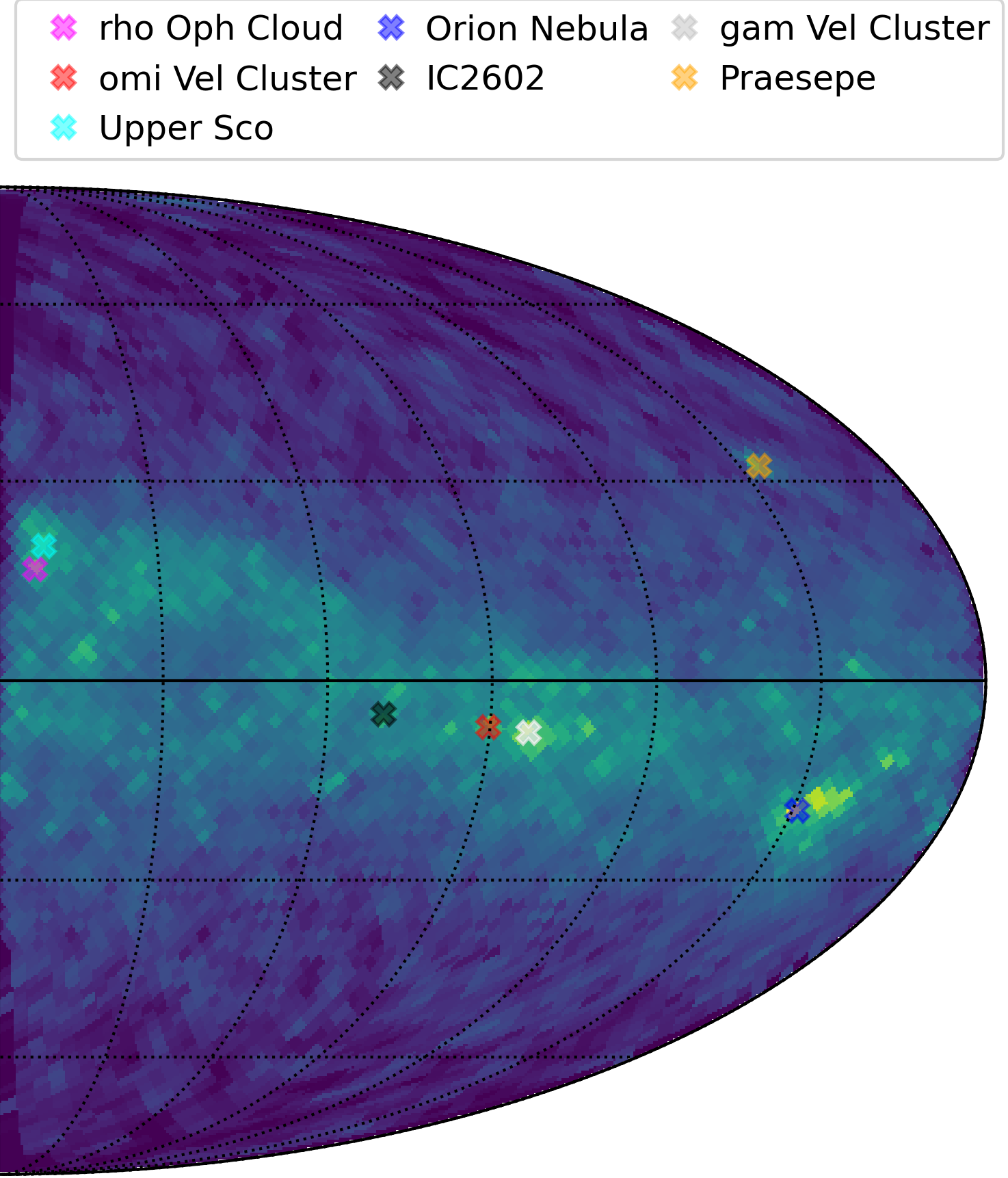}
                \caption{Logarithmic density of the coronal eRASS1 sources with $F_X>5.5\times10^{-14}$~erg\,s$^{-1}$\,cm$^{-2}$. The crosses show the positions of some known stellar clusters.}
                \label{fig: spatial distribution}
        \end{figure}

        Figure~\ref{fig: spatial distribution} shows the density distribution of the coronal eRASS1 sources. We restricted the sample to sources with $F_X>5.5\times10^{-14}$~erg\,s$^{-1}$\,cm$^{-2}$ to reduce the influence of the sensitivity variation for different ecliptic latitudes in eRASS1. The density is enhanced in a region that was formerly described as the Gould belt \citep{gould1879,guillout98a,guillout98b,perrot03} and is nowadays interpreted as a combination of the Radcliffe wave and a structure called the split \citep{alves20}. Many known stellar clusters are visible and a detailed analysis of the stellar clusters in eRASS will be presented by Schneider et al.(in prep.).
        
        \subsection{Completeness and reliability}
        \label{sec: completeness and reliability}
        \begin{figure}[t]
                \includegraphics[width=\hsize]{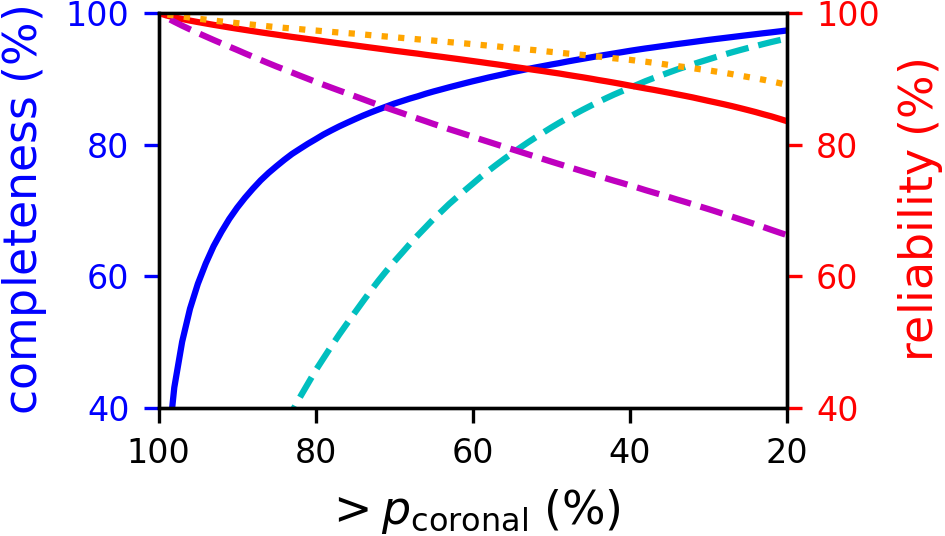}
                \caption{Completeness and reliability of the coronal eRASS1 identifications as a function of the coronal probability. For the dashed cyan and magenta line, only the geometric information is used, while the Bayes map is adopted for the solid blue and red curves. The dotted orange line shows the reliability expected from shifted eRASS1 sources (see text for details).}
                \label{fig: completeness and reliability}
        \end{figure} 
        From the coronal probabilities provided for every eRASS1 source, the number of missed and spurious sources in samples obtained at different probability cutoffs can be estimated as
        \begin{eqnarray}
                N_\mathrm{missed} &=& \sum_i^{N_<} p_{\mathrm{coronal}, i}, \\
                N_\mathrm{spurious} &=& \sum_i^{N_>} (1-p_{\mathrm{coronal}, i}),
                \label{equ: number missed spurious}
        \end{eqnarray}
        and the completeness and reliability can be calculated as follows:
        \begin{eqnarray}
                \mathrm{completeness} &=& \frac{N_> - N_\mathrm{spurious}}{N_> - N_\mathrm{spurious} + N_\mathrm{missed}} \\
                \mathrm{reliabilty} &=& \frac{N_> - N_\mathrm{spurious}}{N_>},
                \label{equ: completeness reliability}
        \end{eqnarray}
        where $N_>$ and $N_<$ are the number of sources above and below the coronal probability cutoff \citep[cf. ][]{freund22}.
        
        Figure~\ref{fig: completeness and reliability} shows the completeness and reliability as a function of the probability cutoff that we obtain if only the geometric properties are considered or additional properties are adopted with the Bayes map. At $p_\mathrm{coronal} \approx 0.53,$ the expected number of coronal sources is recovered and we expect a completeness and reliability of $\sim 91.5~\%$; we note that a value of about 79~\% is reached if only geometric information is applied. If the scientific application requires a sample of higher reliability, the number of spurious associations can be reduced by increasing the coronal probability cutoff. For example, the fraction of chance associations decreases to about 1.4~\% when only sources with $p_\mathrm{coronal} > 0.95$ are considered, as we still obtain a large sample of 83\,400 bona fide coronal X-ray emitters.

        The reliability of the identifications can be independently tested by applying the same procedure to sources at random positions. To conserve the overall source distribution of the eRASS1 catalog, we randomly shifted all eRASS1 sources between 10 and 20~arcmin and applied our HamStar method to these sources. All coronal identifications found for these sources are spurious by construction. However, the number of spurious identifications estimated in this way is too large because the true coronal eRASS1 sources cannot produce a spurious coronal classification. Therefore, the number of coronal associations to shifted sources has to be multiplied by the fraction of noncoronal sources in the eRASS1 catalog ($1 - p_c$). The dotted orange line in Fig.~\ref{fig: completeness and reliability} shows the reliability estimated based on the associations to  shifted sources. This estimation suggests that the reliability obtained by $p_\mathrm{coronal}$ is slightly underestimated (about 3 percentage points at $p_\mathrm{coronal}=0.5$), which might be caused by inaccuracies in the Bayes map; for example, due to under-representation of faint eRASS1 sources.

        \subsection{Comparison with Chandra detections}
        \label{sec: Comparison with Chandra detections}
        \begin{figure}[t]
                \includegraphics[width=\hsize]{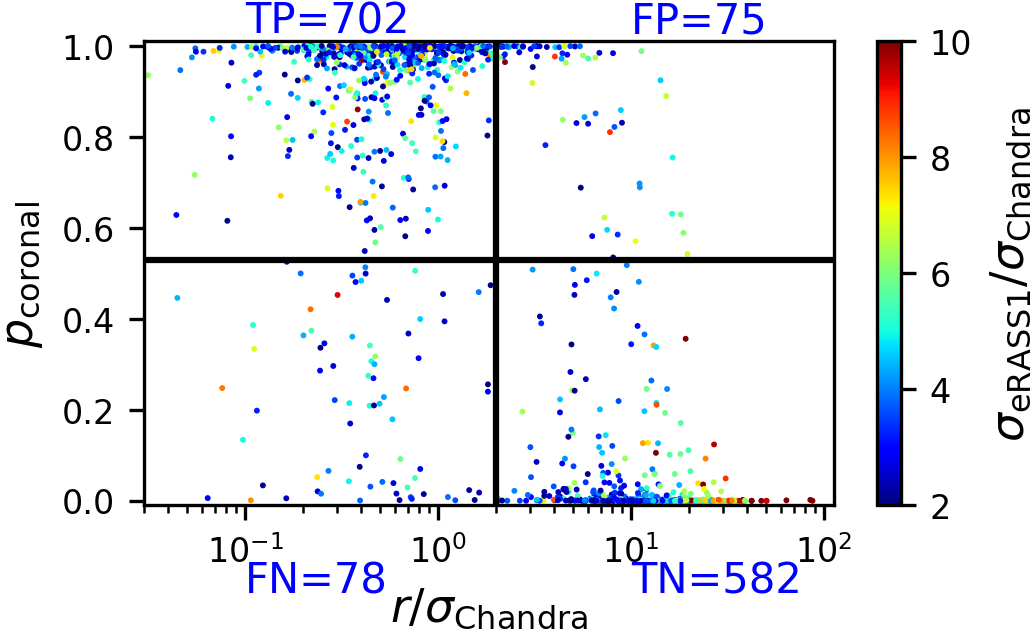}
                \caption{Coronal probability of the eRASS1 sources as a function of the separation between the best coronal counterpart and the \textit{Chandra} detection given in units of positional uncertainty. The solid lines show the thresholds at $p_\mathrm{coronal}=0.53$ and $r/\sigma_\mathrm{Chandra}=2$. The color scales with the ratio of the positional uncertainties from eRASS1 and \textit{Chandra}. The numbers of sources in the resulting four quadrants are specified above and below the figure.}
                \label{fig: eRASS1 Chandra separation p_stellar}
        \end{figure}
        \begin{figure*}[t]
                \centering
                \includegraphics[width=0.89\textwidth]{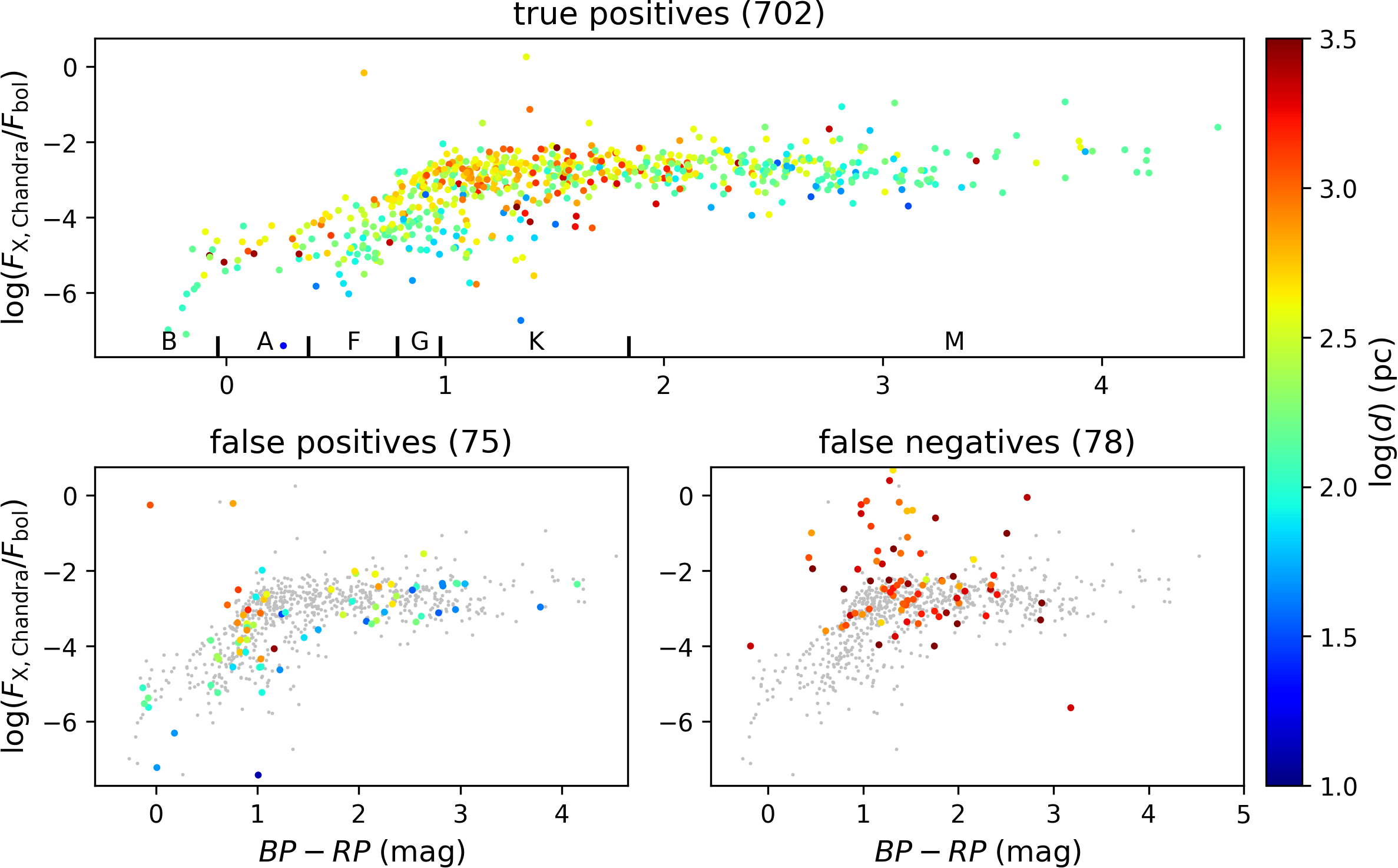}
                \caption{Comparison of the X-ray to bolometric flux ratio as a function of $BP-RP$ color for the eRASS1 sources with a \textit{Chandra} association. The color scales with the distance of the counterparts. The top, left, and right panels show the distribution for the true positives, false positives, and false negatives, respectively. The gray dots show the distribution of the top panel for comparison. }
                \label{fig: comparison eRASS1-Chandra F_X F_bol}
        \end{figure*}
        We subsequently validated our identifications by comparing them with detections from the \textit{Chandra} Source Catalog 2.0 \citep[CSC;][]{CSC}. To minimize the number of spurious identifications, we considered only eRASS1--\textit{Chandra} associations within 20~arcsec and $2\sigma$ of the combined positional uncertainties from the eRASS1 and \textit{Chandra} source. We excluded \textit{Chandra} detections without flux measurement in the ACIS broad and HRC wide band. Furthermore, we considered only unique \textit{Chandra} counterparts and only sources for which the positional accuracy of \textit{Chandra} is at least a factor of two better than for the eRASS1 source. 
        
        In this fashion, we obtained 3\,515 eRASS1 sources with a \textit{Chandra} association fulfilling our criteria and 1\,437 of them have an eligible coronal counterpart within our matching radius. For these sources, we show in Fig.~\ref{fig: eRASS1 Chandra separation p_stellar} the coronal probability as a function of the angular separation between our best counterpart and the \textit{Chandra} detection divided through the \textit{Chandra} positional uncertainty $\sigma_\mathrm{Chandra}$. In the following, we consider an eRASS1 source with $p_\mathrm{coronal}>0.53$ as a coronal identification and we assume this identification to be confirmed if the separation of the best counterpart to the \textit{Chandra} detection is $r<2\sigma_\mathrm{Chandra}$. The majority of our coronal identifications are indeed confirmed by \textit{Chandra} (true positives). The number of associations rejected by \textit{Chandra} (false positives) is formally larger than expected from our reliability estimation, but we emphasize that about 13.5~\% of the true counterparts are located outside the $2\sigma$ radius in a Rayleigh distribution. Also, the number of counterparts confirmed by \textit{Chandra} but with a low coronal probability (false negatives) is slightly larger than expected. 
        
        In Fig.~\ref{fig: comparison eRASS1-Chandra F_X F_bol} we compare the distributions of the X-ray\footnote{We adopted the \textit{Chandra} X-ray flux in the ACIS broad band, if available, and otherwise in the HRC wide band estimated for an APEC model and converted these fluxes into the 0.2--2.3~keV \textit{eROSITA} energy band.} to bolometric flux ratio versus color for the true positives, false positives, and false negatives. Our coronal identifications rejected by \textit{Chandra} are similar to the confirmed counterparts. Many of the counterparts with a low coronal probability despite a small separation to the \textit{Chandra} detection have large fractional X-ray fluxes and 19 of them are classified in the Simbad database \citep{Simbad} as accreting or extragalactic objects with a spurious parallax in \textit{Gaia} DR3. As our algorithm is trained for coronal emitters, the probability of these sources is strongly downweighted, which explains why the number of false negatives is larger than expected. There is also a small number of sources with high fractional X-ray fluxes that are identified by HamStar, and these sources are flagged as noncoronal (see Sect.~\ref{sec: non-coronal contamination}). There are also confirmed associations with a small probability that are located near or below the saturation limit, and these sources seem to be true coronal emitters that are missed by our identification procedure. These counterparts have larger distances than the sources with a high coronal probability, and, furthermore, they are typically X-ray faint and have a low detection likelihood. This may indicate that our sample of coronal eRASS1 sources is missing some true identifications with these properties.
        
        We note that 262 of the sources in Fig.~\ref{fig: eRASS1 Chandra separation p_stellar} and 146 of the true positives are extended \textit{Chandra} sources, although the extent is, for most sources, $<5$~arcsec and probably not always reliable. One possible explanation is that these sources are close binaries whose X-ray emission is inconsistent with a point-like source but cannot be resolved by \textit{Chandra}. Indeed, about one-third of the \textit{Gaia} counterparts have a \texttt{RUWE} $<1.4$, which might indicate the existence of a binary component.
        
        \subsection{Comparison with identifications from NWAY}
        \label{sec: comparison with identifications from NWAY}
        We next compared our identification with the association presented in Salvato et al. (in prep.), which also uses \textit{Gaia} DR3 as one of the multiwavelength catalogs.  The authors, in addition to the separation between primary (\textit{eROSITA}) and secondary (\textit{Gaia}) catalogs and respective positional uncertainties  and number density, also use a prior defined based on about 24\,000  training sources detected by \textit{XMM-Newton} and \textit{Chandra} and with secure counterparts. The methodology is almost the same as that presented in \citet{salvato18} for eFEDS. The major changes are the correction of the source position for proper motion and the definition of the prior.
        Specifically, Salvato et al. (in prep.) consider photometry, colors, signal-to-noise ratio (S/N), and proper motion. As opposed to this paper, which focuses on the identification of coronal stars, the goal of Salvato et al. (in prep.) is to determine all types of possible X-ray emitters. However, Salvato et al. (in prep.)  already pointed out that the identifications obtained with \textit{Gaia} DR3 are less complete and pure than in the LS10 area (about 73\% in completeness and purity instead of about 95\%) because the number of features available in \textit{Gaia} DR3 is limited.
For our comparison, we excluded  the sources located more than 1.5~mag below the main sequence (cf. Sect.~\ref{sec: color-magnitude diagram}).
        
        In Table.~\ref{tab: NWAY comparison p_any}, we compare the eRASS1 identifications from NWAY and HamStar. As HamStar only identifies coronal sources, there are 94\,200 eRASS1 sources with $p_\mathrm{coronal}<0.53$ that are identified by NWAY with a different counterpart and $p_\mathrm{any}>0.02$. For this reason, the NWAY counterpart is likely extragalatic, and therefore confirms that the coronal HamStar counterpart is probably not the correct association. There are 137\,500 of our good coronal eRASS1 sources (with $p_\mathrm{coronal}>0.53$) that also have a counterpart in NWAY, and for 123\,700 of them the same counterpart is also classified as a good identification by NWAY ($p_\mathrm{any}>0.02$). This agreement of 90.0~\% is slightly lower than the expected reliability of HamStar but is larger than the purity of the NWAY identifications, and is therefore within the expected range. For an additional 3\,100 sources, NWAY finds the same \textit{Gaia} DR3 source as the best counterpart but provides a low probability that this source is the correct association of the eRASS1 source, while 
        HamStar produces a high probability. On the other hand, 17\,800 sources have a high $p_\mathrm{any}$ in NWAY, while according to HamStar this counterpart is more likely not a coronal emitter. 
        \begin{table*}[t]
                \caption{Comparison of the number of sources identified with NWAY and HamStar for sources with the same and different counterparts (rounded to the nearest hundred)}
                \label{tab: NWAY comparison p_any}
                \centering
                \begin{tabular}{lllll}
                        \hline
                         & \multicolumn{2}{c}{same counterparts} & \multicolumn{2}{c}{diff counterparts} \\
                         & $p_\mathrm{coronal} < 0.53$ & $p_\mathrm{coronal} > 0.53$ & $p_\mathrm{coronal} < 0.53$ & $p_\mathrm{coronal} > 0.53$ \\
                        \hline \hline
                        $p_\mathrm{any} > 0.02$ & 17\,800 & 123\,700 & 94\,200 & 8\,900 \\
                        $p_\mathrm{any} < 0.02$ & 50\,600 & 3\,100 & 83\,300 & 1\,800 \\
                        \hline
                \end{tabular}
        \end{table*}

\begin{figure*}[t]
                \centering
                \includegraphics[width=\hsize]{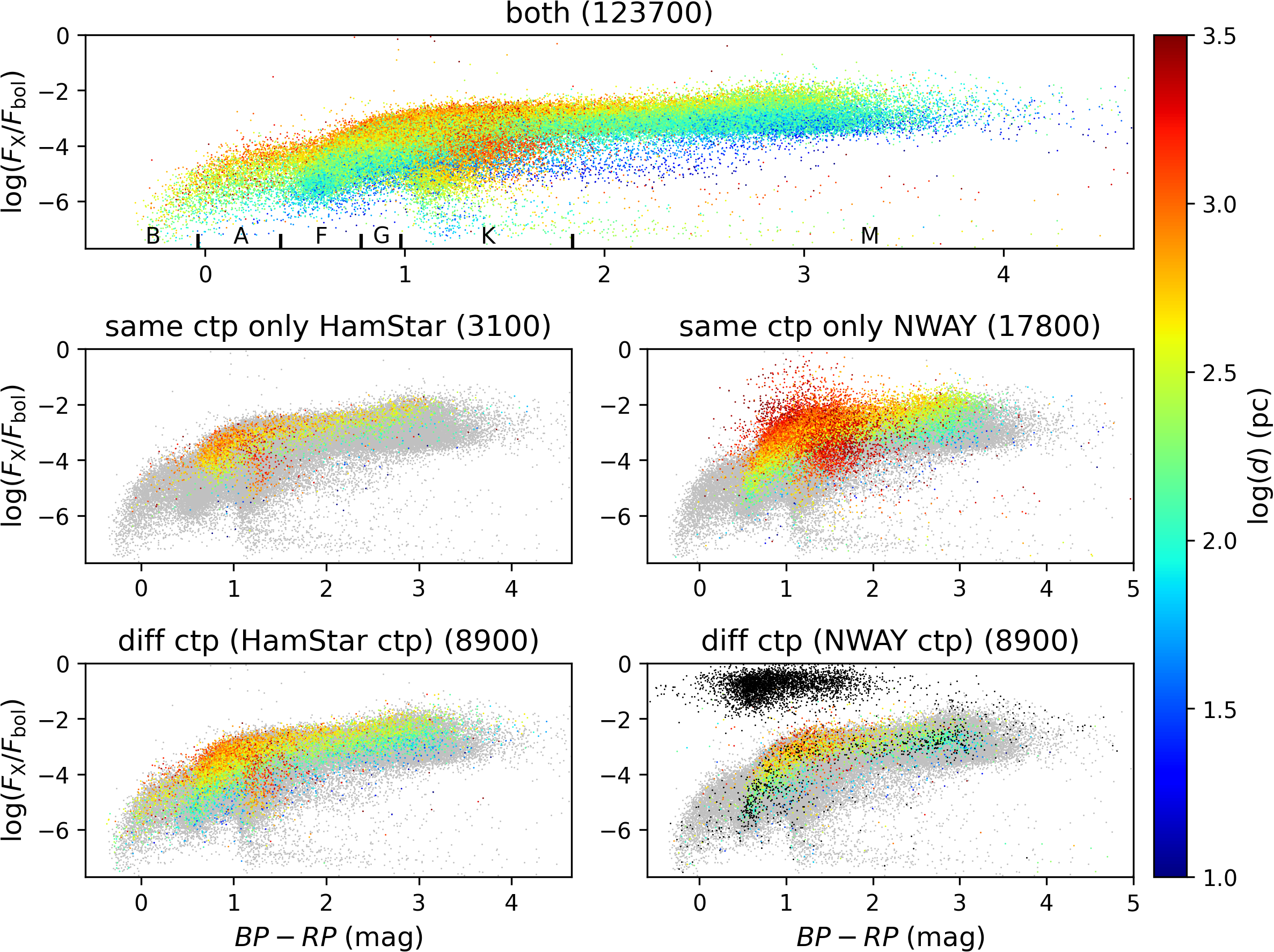}
                \caption{Comparison of the X-ray to bolometric flux ratio as a function of the $BP-RP$ color for the eRASS1 sources with the same and different counterparts by NWAY and our method. The top panel shows the distribution of the consistently identified sources, and the middle panel shows the sources only identified by our method (left) and NWAY (right). In the bottom panels, the sources with different counterparts are compared, while the best counterparts from our method and NWAY are shown in the left and right panels, respectively. The numbers of sources in the panels are specified in parentheses (cf. Table~\ref{tab: NWAY comparison p_any}). The color scales with the distance of the counterparts and ineligible coronal counterparts from NWAY are shown as black dots. }
                \label{fig: NWAY comparison F_X F_bol}
        \end{figure*}

        In Fig.~\ref{fig: NWAY comparison F_X F_bol} we compare the X-ray to bolometric flux ratio with color for these sources. The properties of the sources with a high $p_\mathrm{coronal}$ and a low $p_\mathrm{any}$ are very similar to those of the consistently identified sources. A few of the sources that are identified by NWAY but rejected by HamStar have reasonable properties and these sources might be missed by our procedure; however, many of these sources have much larger distances and fractional X-ray fluxes. These latter are unlikely to be coronal emitters but some might be the correct association of other Galactic X-ray sources, such as cataclysmic variables (CVs). 

        For 8\,900 eRASS1 sources, $p_\mathrm{any}$ and $p_\mathrm{coronal}$ are high, but the two methods find different best counterparts. In the bottom panel of Fig.~\ref{fig: NWAY comparison F_X F_bol}, we compare the properties of the best counterparts from NWAY and HamStar. The counterparts that we identify as best associations have similar properties to the consistent identifications. In 3\,400 of these cases, the best NWAY counterpart is an eligible coronal candidate that is also considered by HamStar. Both counterparts also often have a similar distance, which means that they likely belong to the same binary system or stellar cluster (cf. Sect.~\ref{sec: multiplicity}) and the best NWAY counterpart also has properties typical of coronal X-ray emitters. Some of the best NWAY counterparts that are not considered by HamStar also have similar properties to coronal sources. These counterparts typically miss a significant parallax in \textit{Gaia} DR3. Most of the best NWAY counterparts not considered by HamStar have much larger fractional X-ray fluxes and these are probably extragalactic sources. These eRASS1 sources therefore have both a reasonable coronal and extragalactic counterpart. Applying the same criteria as in Sect.~\ref{sec: Comparison with Chandra detections}, we find a Chandra counterpart for 75 of the eRASS1 sources with different best counterparts. For 10 sources, the separation between the \textit{Chandra} detection and the HamStar counterpart is within $2\sigma$ of the \textit{Chandra} positional accuracy, but the NWAY counterpart has a larger separation and in 21 cases, the NWAY but not the HamStar counterpart is confirmed by the \textit{Chandra} detection. However, the \textit{Chandra} counterpart is consistent with both identifications from HamStar and NWAY for 39 sources, because either one \textit{Chandra} source is located close to both identifications (19 cases) or there are two \textit{Chandra} detections in the vicinity of the eRASS1 source, one confirming the HamStar counterpart and the other being consistent with the NWAY identification (20 cases). As the separation between the two counterparts is smaller than the angular resolution of \textit{eROSITA}, it is likely that the eRASS1 source is a superposition of the emission from the two. For 5 sources, the \textit{Chandra} counterpart is neither
consistent with the HamStar nor the NWAY identification.
        
        In summary, we conclude that the identifications generally agree reasonably well between NWAY and HamStar. Sources with different counterparts often have low detection likelihoods, and therefore some of them might be spurious detections. In addition, the positional uncertainties are generally large and both counterparts usually have large angular separations to the nominal \textit{eROSITA} source position. In such cases, the weighting of the physical properties becomes more important relative to the geometry, and so it is expected that the identifications differ more strongly for such sources. For a small number of sources with different counterparts, Chandra detections with a better positional accuracy are available. For most them, the Chandra position is consistent with the counterparts from NWAY and HamStar, and for some sources, both counterparts seem to contribute to the X-ray emission that cannot be resolved by \textit{eROSITA}. Due to the small sample sizes, we can only note that the identifications of both methods are similarly good for sources with different best counterparts, and we further emphasize that, for 88~\% of the coronal sources, NWAY and HamStar agree on the counterpart. Overall, the specialized HamStar method provides better results for the identification of coronal sources, because it achieves a higher completeness and reliability than NWAY (91.5~\% compared to 73~\%). Furthermore, some NWAY identifications have clearly noncoronal properties. Although they might be the correct association of a CV, these objects are more difficult to distinguish from a coronal sample than with the HamStar catalog. Nevertheless, the NWAY identifications are useful to coronal science, because they can inform us about potential alternative extragalactic counterparts not considered by HamStar. We therefore flagged sources with an alternative counterpart in NWAY.
        
        \subsection{Catalog of the coronal identifications}
        \label{sec: publication of the stellar identifications}
        We created two catalogs of our coronal eRASS1 identifications. The first catalog contains the 139\,700 likely coronal eRASS1 sources (i.e., $p_\mathrm{coronal}>0.53$) with all counterparts with $p_{ij} > 0.1$. These are the sources discussed in the present paper. Some of these sources have multiple counterparts and therefore contain multiple entries in the catalog. Additionally, we provide possible coronal eRASS1 sources with $0.53 > p_\mathrm{coronal} > 0.2$ and $p_{ij} > 0.1$. We expect most of these counterparts to be spurious associations, but this sample still contains some correct coronal X-ray sources. Both catalogs are available via CDS/VizieR and via the \textit{eROSITA} DR1 website\footnote{\url{https://erosita.mpe.mpg.de/dr1/AllSkySurveyData_dr1/Catalogues_dr1/}} and we describe the columns of the catalog in Appendix~\ref{sec: column description}.

        \section{Properties of the coronal eRASS1 sources}
        \label{sec: Properties of the stellar eRAS1 sources}
        With the HamStar method, we identify 149\,300 likely counterparts to 139\,700 eRASS1 sources and we describe the properties of these sources in the following sections. As described in Sect.~\ref{sec: non-coronal contamination}, some of these sources have properties or Simbad classifications that are not typical of coronal sources and we flag 1\,161 as likely noncoronal emitters. This leaves us with 138\,800 coronal eRASS1 sources whose best counterpart is not flagged. Furthermore, we find more than one plausible counterpart for some eRASS1 sources, and  we generally cannot estimate the contribution of the individual counterparts to the X-ray emission as discussed in Sect.~\ref{sec: multiplicity}. When restricting the sample to coronal eRASS1 sources with one counterpart with a matching probability of $p_{ij}>0.5$, we obtain 137\,300 coronal eRASS1 sources; these are analyzed in the following sections.
        
        \subsection{Noncoronal contamination}
        \label{sec: non-coronal contamination}
        Although our method is optimized for coronal identifications, some of our sources have properties that are not typical of coronal X-ray emitters. The properties of these counterparts typically strongly overlap with those of random associations, and therefore the fraction of chance alignments is relatively large. Nevertheless, some of these counterparts are likely the correct identification of the eRASS1 source but their X-ray emission is probably caused by compact objects possibly through an accretion process or by an extragalactic object with a spurious parallax in \textit{Gaia} DR3. Specifically, we flagged 1\,096 of our likely counterparts with $\log(F_X/F_\mathrm{bol})>-1$, $L_X>10^{32}$~erg~s$^{-1}$, or located more than 1.5~mag below the main sequence.
        
        To find further counterparts of noncoronal source types, we cross-matched all of our counterparts with the Simbad database and obtained a Simbad classification for 68\,100 of the 149\,300 counterparts in our catalog of likely associations. Of our flagged counterparts, 285 and 26 sources are known in Simbad as accreting and extragalactic objects, respectively. The Simbad classifications reveal a further 58 and 7 known accreting and extragalactic sources, respectively, not identified by our filter criteria and we flagged these sources too. Furthermore, we specify the Simbad name and classification for all counterparts in our catalog if available.
        
        \subsection{Multiplicity}
        \label{sec: multiplicity}
        \begin{figure}[t]
                \centering
                \includegraphics[width=0.8\hsize]{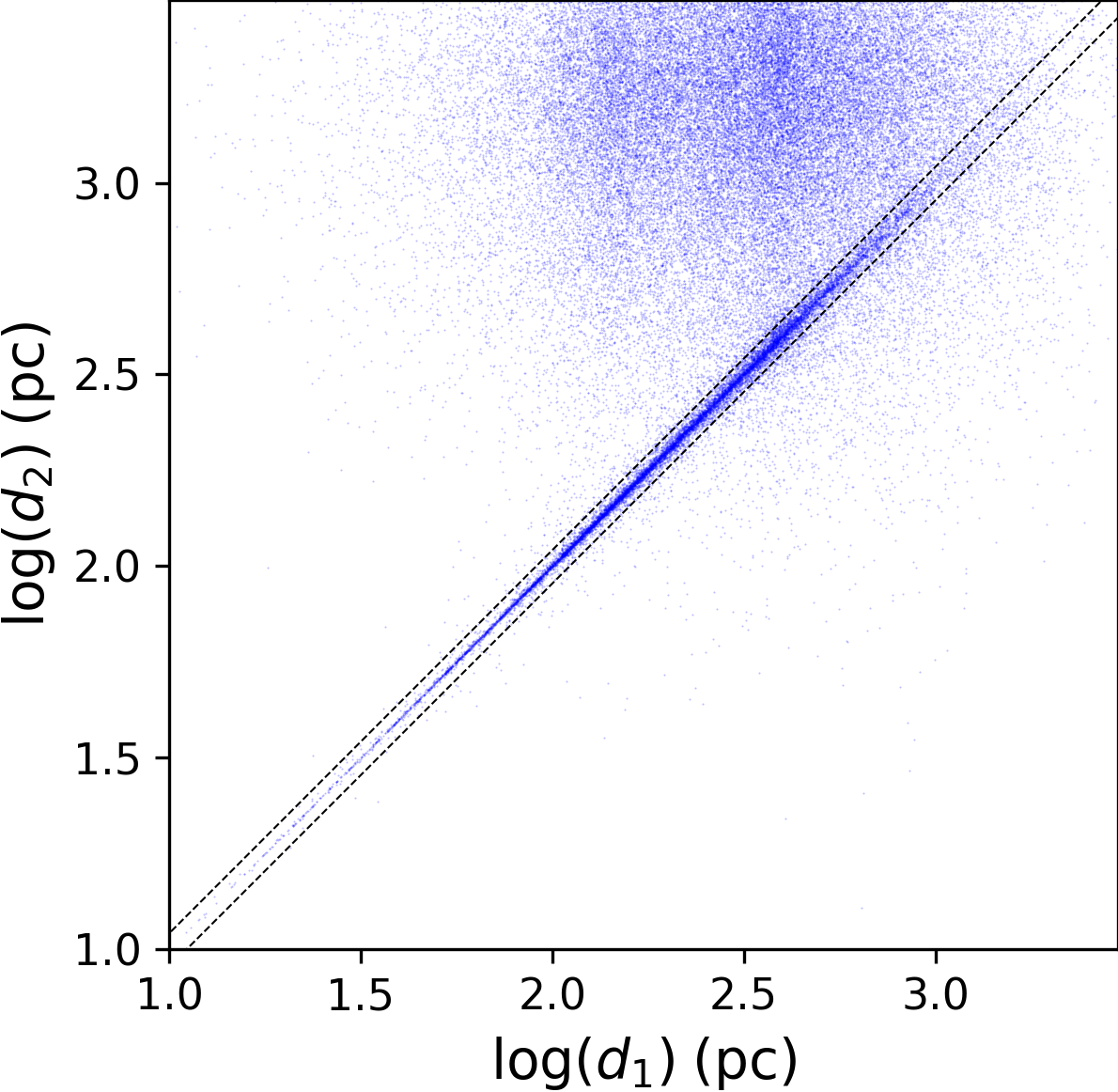}
                \caption{Comparison of the distances of the best and second-best counterparts for the coronal eRASS1 sources with multiple counterparts. For sources located between the dashed lines, the distances do not differ by more than 10\%. }
                \label{fig: Multiplicity distance comparison eRASS1}
        \end{figure}
        \begin{figure}[t]
                \centering
                \includegraphics[width=\hsize]{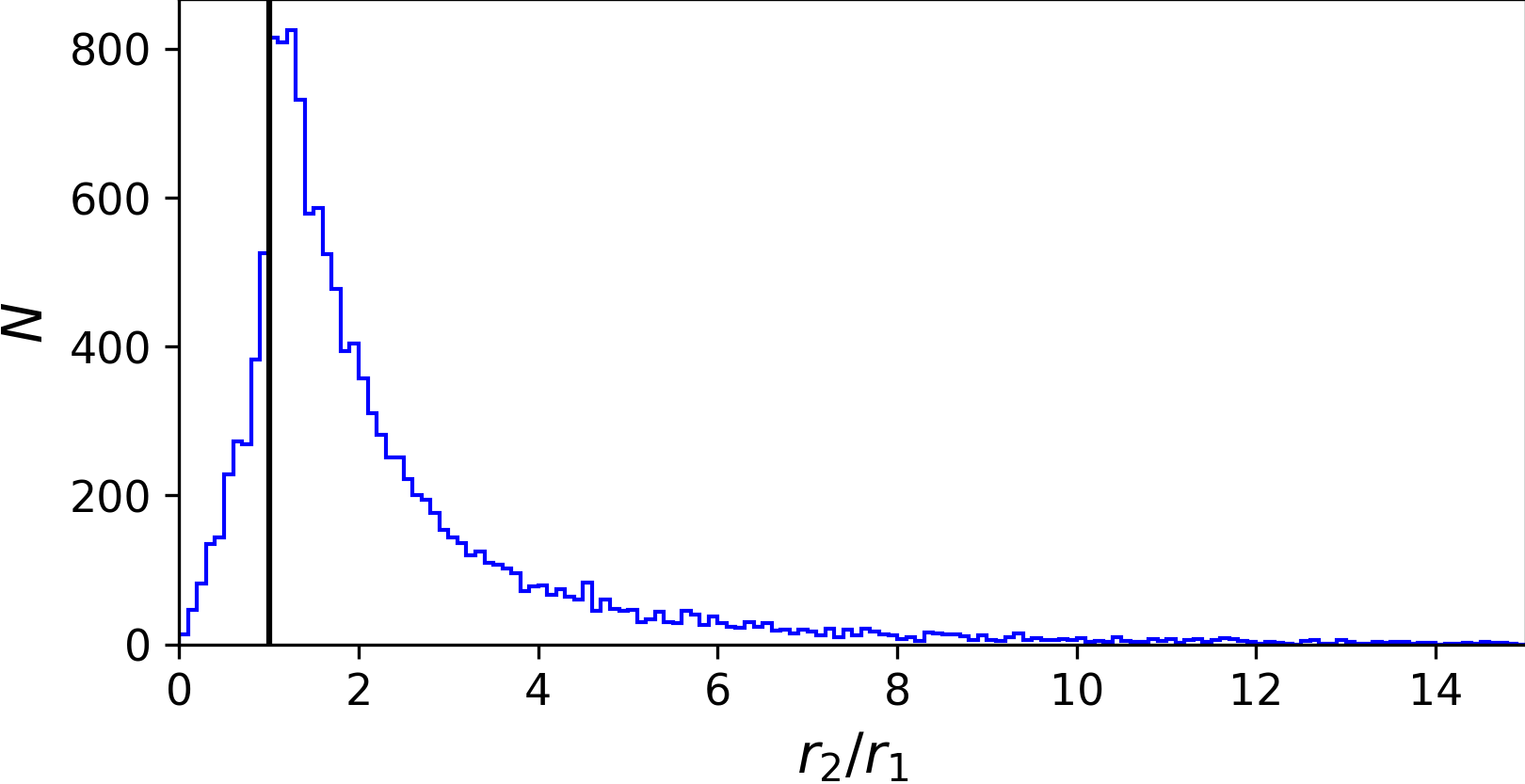}
                \caption{Distribution of the ratio of the angular separations between the eRASS1 source and the second-best and best counterparts. On the vertical line, both counterparts have the same separation.}
                \label{fig: Multiplicity angular separations comparison eRASS1}
        \end{figure}
        For 70\,600 of the 138\,800 coronal eRASS1 sources, we find multiple counterparts within our $5\sigma$ search radius. In Fig.~\ref{fig: Multiplicity distance comparison eRASS1} we compare the distances of the best and second-best counterparts for these sources. For most sources, these distances are uncorrelated with the typically much larger distances of the second-best counterparts, but there is a narrow band of sources with similar parallaxes up to distances of about 1~kpc. Specifically, for 13\,600 of 130\,600 coronal eRASS1 sources within 1~kpc, the distances of the best and second-best counterparts do not differ by more than 10~\%. Most of these counterparts also have similar proper motions, although the difference is often larger than the uncertainties given in \textit{Gaia} DR3. The counterparts are therefore correlated comoving pairs and are probable companions in a wide binary system or belong to the same stellar cluster. The angular separation between the counterparts is
        typically smaller than 10~arcsec and peaks near the \textit{Gaia} resolution limit of about 2~arcsec. Figure~\ref{fig: Multiplicity angular separations comparison eRASS1} shows the ratio of the angular separations between the eRASS1 source and the two counterparts. The best counterpart is in most cases the match with the smallest separation, but the distribution peaks at separations  similar to that of both counterparts. This favors the argument that, for many of these sources, both counterparts contribute approximately equally to the X-ray emission, but the source is unresolved by \textit{eROSITA}. However, for a specific source, we cannot quantify the contribution of the individual counterparts. In our catalog, the highest matching probabilities are typically found for  sources that are earlier in type, brighter, and have the smallest angular separation to the eRASS1 source. In the following, we analyze the properties of those counterparts with a matching probability of $p_{ij}>0.5$ but we note that part of the X-ray emission might be produced by a companion.

        \begin{figure}[t]
                \includegraphics[width=\hsize]{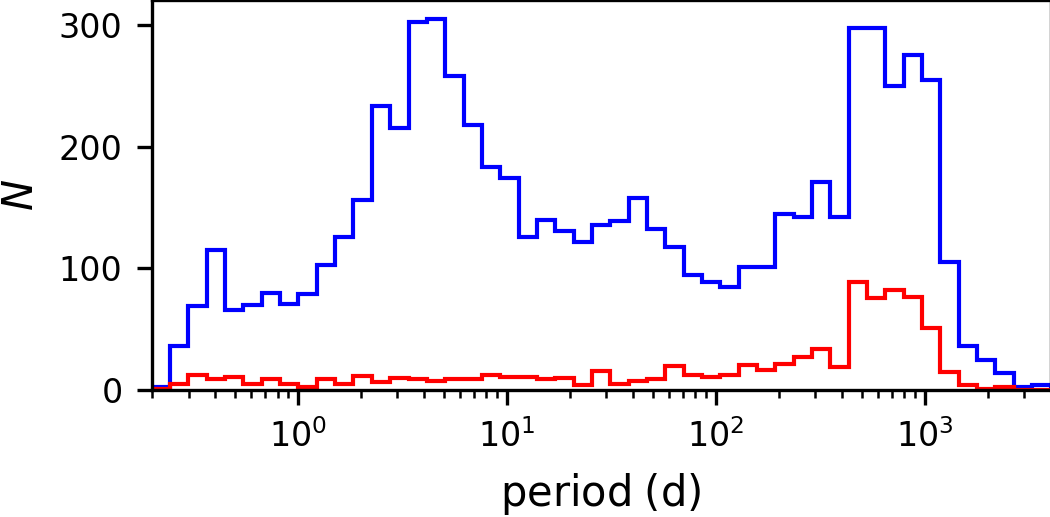}
                \caption{Distribution of the binary periods for the coronal eRASS1 sources (blue) and the counterparts to the shifted sources (red)}
                \label{fig: Multiplicity binary periods eRASS1}
        \end{figure}
        In \textit{Gaia} DR3, 6\,699 of our coronal eRASS1 sources are identified as nonsingle stars unresolved by \textit{Gaia} based on astrometric, photometric, and spectroscopic measurements and combinations thereof \citep{halbwachs23,holl23,mowlavi23,arenou23}. The completeness of the nonsingle star identification is not well defined and depends on many parameters, such as the brightness of the source or the orientation of the orbit, and the dependence also differs depending on the identification technique. The fraction of unresolved binaries is much smaller for the shifted eRASS1 sources and only 830 of 304\,000 counterparts to the shifted sources are identified as nonsingle stars. Although these 304\,000 counterparts were not filtered for being coronal, it is obvious that the number of binaries is much larger in the sample of real counterparts than among random counterparts; in other words, the HamStar stars are more frequently binaries. In Fig.~\ref{fig: Multiplicity binary periods eRASS1}, we compare the binary periods of the counterparts to the true and shifted eRASS1 sources. Our sample of X-ray-emitting sources contains a particularly high number of short-period binaries with periods of between about 2 and 10~d. These sources are known to preserve a high X-ray activity for a longer time due to tidal interaction \citep[e.g., ][]{zahn77}. 
        
        \begin{figure*}[t]
                \centering
                \includegraphics[width=0.89\textwidth]{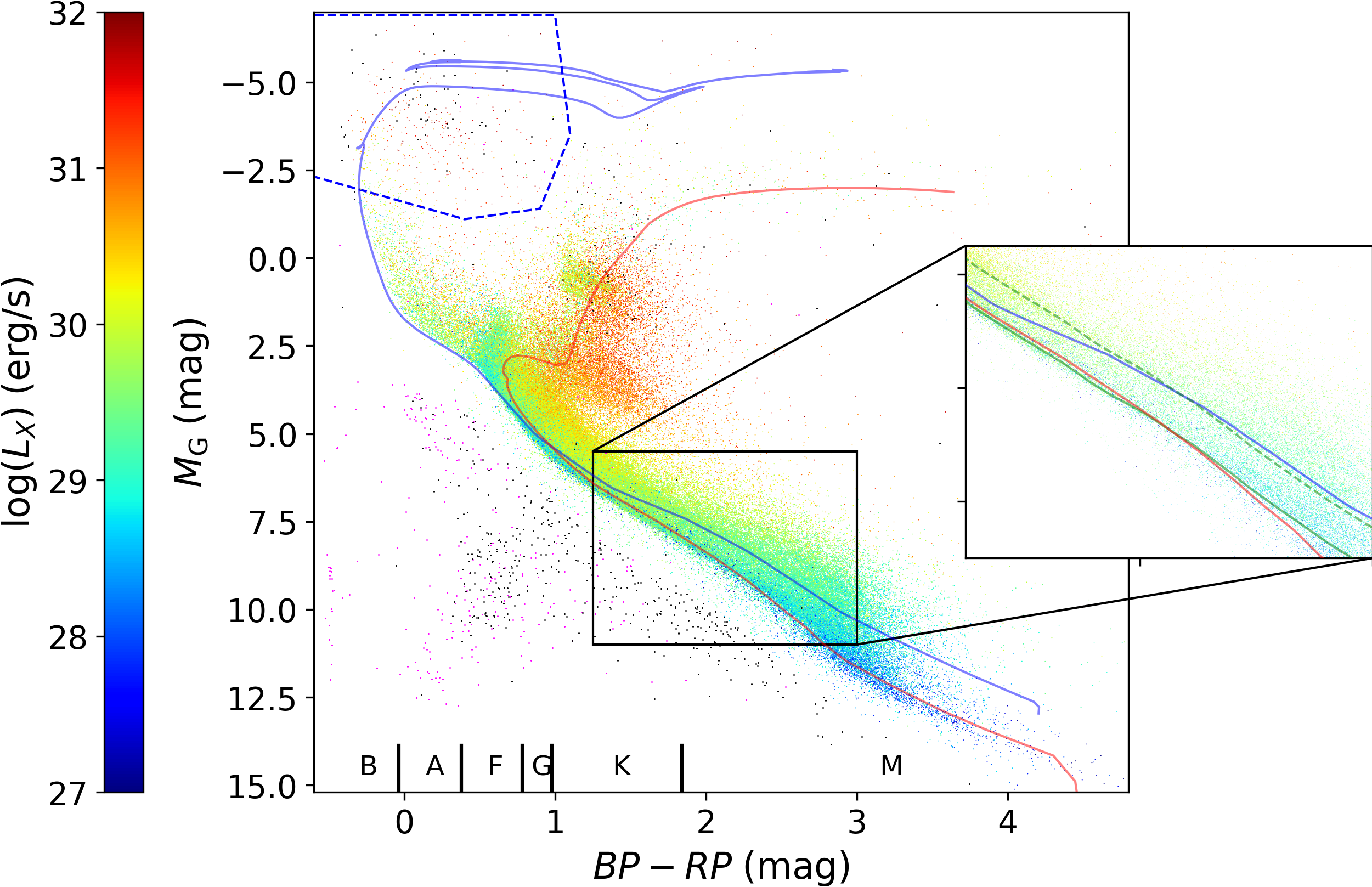}
                \caption{Color--magnitude diagram of the coronal eRASS1 sources. The color scales with X-ray luminosity. Magenta and black sources are flagged as noncoronal emitters; the magenta sources are known accreting or extragalactic objects in the Simbad database. Sources in the area outlined by the dashed blue line are flagged as \texttt{OB\_STAR}. The blue and red lines show the PARSEC isochrones for stellar ages of $4\times10^7$ and $4\times10^9$~yr \citep{bressan12}, respectively, while the solid and dashed green lines in the inset show the $1.3\times10^8$~yr isochrone for single and binary stars.}
                \label{fig: CMD eRASS1}
        \end{figure*}
        \subsection{Color--magnitude diagram}
        \label{sec: color-magnitude diagram}
        Figure~\ref{fig: CMD eRASS1} shows the color--magnitude diagram (CMD) of the coronal eRASS1 sources, with the black and magenta dots representing the sources flagged as noncoronal. This population is similar,  overall, to the coronal RASS sources presented by \citet{freund22} and covers all spectral types from early B- and O-type stars (shifted to the right due to reddening) down to the bottom of the main sequence. Furthermore, because of the much larger sample size, the edges of the distribution are   also well populated. For example, the number of sources decreases for stars redder than $BP-RP\approx2.8$~mag (corresponding to spectral type M3.5), as is the case for the coronal RASS sources, but with eRASS1, we still find 1\,331 stars later than $BP-RP=3.4$~mag (M5). We emphasize that the X-ray emission of early-type stars is not produced in a corona as in stars of later spectral types. Nevertheless, early-type stars do not appear as a separate population in our plots (compare also Sect.~\ref{sec: X-ray over bolometric flux}). Due to their large distances, early-type stars are often shifted to the right in the CMD because of reddening and we flagged 429 sources in the region outlined by the dashed blue line in Fig.~\ref{fig: CMD eRASS1} as likely reddened \texttt{OB\_STAR} in the catalog. Many of these sources indeed have a large reddening value in \textit{Gaia}~DR3, but this measurement is not available for all of our counterparts. The binary sequence located 0.75~mag above the main sequence is clearly visible and there are further, presumably young late-type stars above the main sequence. Indeed, about 7\,500 and 9\,000 of our counterparts are known binaries and young stellar objects or candidates, respectively, in the Simbad database. 

        The sample of coronal eRASS1 sources also contains evolved stars and 17\,500 are more than 2.2~mag brighter than the main sequence with a color of $0.7<BP-RP<2$~mag. These giants and subgiants often exhibit high X-ray luminosity and are probably associated with active binaries like the RS CVn-type systems. The medium-luminosity giants are located more to the left-hand side of the giant branch. We also find some M-giants beyond the dividing line \citep{linsky79, huensch96} associated with an eRASS1 source and these objects are discussed in detail by Schmitt et al.(in prep.). Many of the M-giants, as well as some other very luminous giants, are flagged as noncoronal sources.
        
        Some of our identifications are located below the main sequence. The X-ray emission of these sources is likely produced by a compact object often through accretion processes and not in a corona. We flagged 666 sources located more than 1.5~mag below the main sequence as noncoronal emitters (see Schwope et al.(in prep.) for a detailed identification and discussion of these source types). Many of the bluer objects are known in Simbad as being noncoronal emitters but most of the flagged sources located in the K and M regimes do not have a noncoronal Simbad classification. For some of these sources, this location in the CMD might  be caused by an erroneous parallax or photometry. Nevertheless, the inferred properties are not coronal and therefore the sources are flagged.
        
        \subsection{X-ray over bolometric flux}
        \label{sec: X-ray over bolometric flux}
\begin{figure*}[t]
                \centering
                \includegraphics[width=0.89\textwidth]{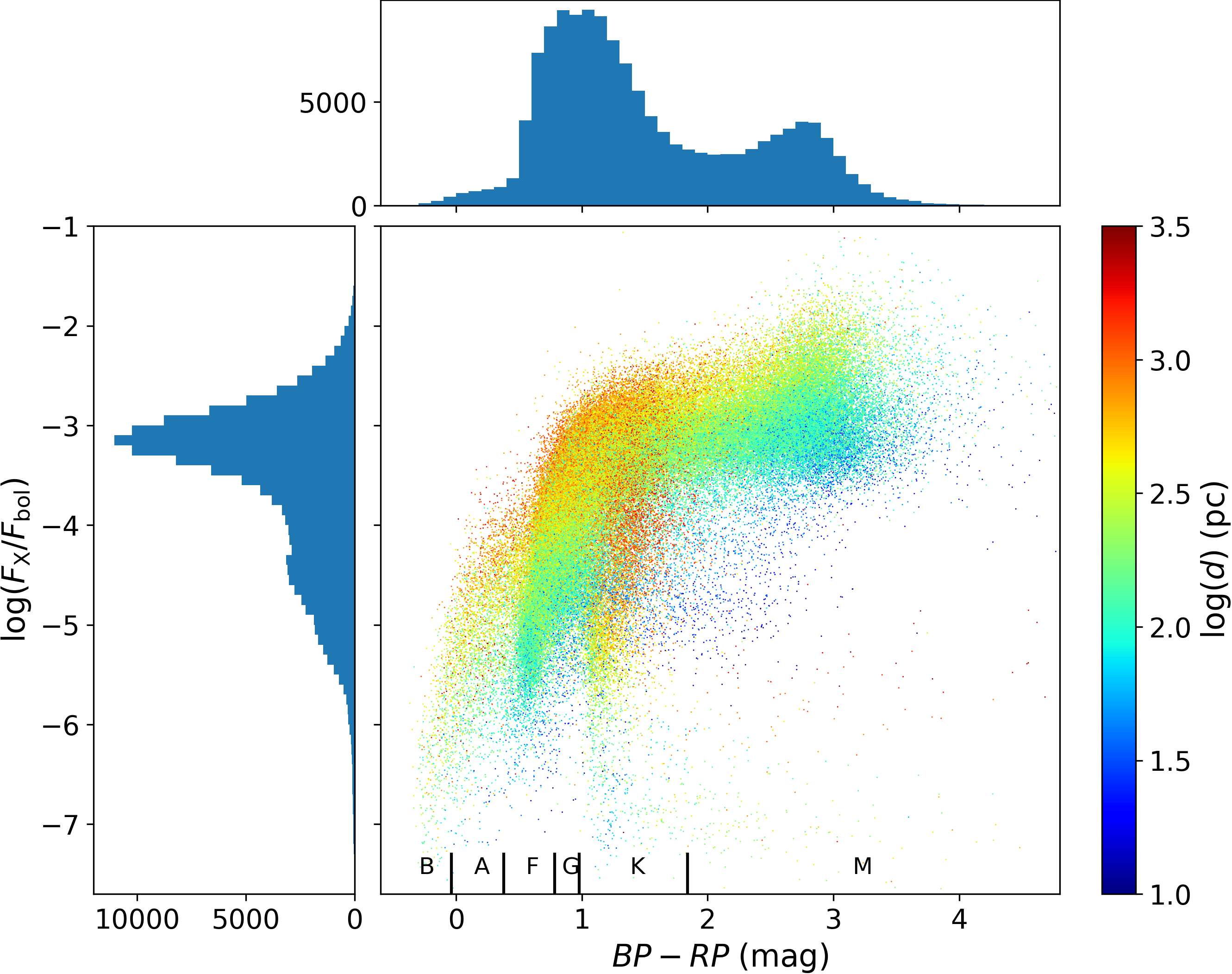}
                \caption{X-ray to bolometric flux ratio as a function of $BP-RP$ color for the coronal eRASS1 sources. The color scales with the distance of the counterpart and the histograms in the upper and left panels show the distribution of the $BP-RP$ color and that of the fractional X-ray flux, respectively. }
                \label{fig: F_X/F_bol eRASS1}
        \end{figure*}

        Figure~\ref{fig: F_X/F_bol eRASS1} shows the fractional X-ray fluxes versus color for the coronal eRASS1 sources that are not flagged for the noncoronal properties or Simbad classifications. Their distribution is generally similar to that of the coronal RASS sources \citep{freund22} and the onset of convection for early F-type sources and the saturation limit at $\log(F_X/F_\mathrm{bol})=-3$ are clearly visible. The early-type stars at low activity levels partly overlap with binaries containing an X-ray-faint A-type star and a companion of a later spectral type emitting coronal X-rays. In such binaries, the optical emission is dominated by the A-type star and the low-mass companion dominates the X-ray emission. As for the coronal RASS sources, those above the saturation limit are likely detected during a flare, but the reliability of these identifications is lower as indicated by the generally lower coronal probabilities. Due to the higher sensitivity of eRASS1, sources  are detected down to lower fractional X-ray fluxes and later spectral types compared to RASS. Some X-ray detections of optically very bright sources are probably caused by optical loading and are flagged in the eRASS1 catalog.
        
        The distances of the counterparts generally increase with the fractional X-ray flux because highly active sources are rarer but can be detected in a larger volume. 
        The giant branch is located at spectral types G and early K and is visible in Fig.~\ref{fig: F_X/F_bol eRASS1} as the sources with larger distances. The giants detected in eRASS1 have activity levels in the range of roughly $-4 < \log(F_X/F_\mathrm{bol}) < -7$ and the saturation limit is only reached for subgiants. Most of the M-giants visible in Fig.~\ref{fig: CMD eRASS1} have quite a low fractional X-ray flux. 
        
        
        \subsection{Rotation--activity relation} 
\begin{figure}[t]
                \includegraphics[width=\hsize]{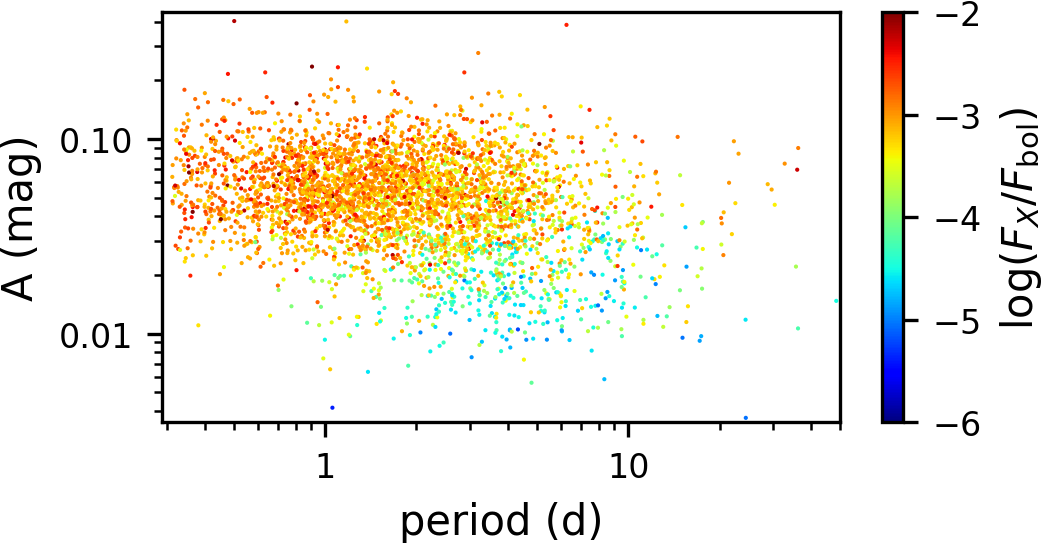}
                \caption{Rotation periods as a function of the amplitude of the modulation for the coronal eRASS1 sources with rotation modulation from \citet{distefano23}. The color scales with the ratio of X-ray to bolometric flux.}
                \label{fig: period-amplitude eRASS1}
        \end{figure}

        For 3\,721 of the coronal eRASS1 sources, we find a measurement of the rotation modulation from \citet{distefano23}. This is the largest sample of sources with measured X-ray flux and rotation period obtained so far. Due to the complex \textit{Gaia} scanning law, the completeness of the rotation periods strongly varies over the sky. According to \citet{distefano23}, the detection of rotation periods of $<5$~d is favored and about 70\%\ to 80\% of the detected periods are expected to be correct. Figure~\ref{fig: period-amplitude eRASS1} shows the period--amplitude diagram of the coronal eRASS1 sources. Most of these sources have rotation periods of between 5d and less than 1d and the amplitudes and rotation periods are well correlated with the X-ray to bolometric flux ratio. Compared to the period--amplitude diagram of \citet{distefano23} (shown in their Fig.~21), the population of fast rotators ($P<2$~d) with low amplitudes ($A<0.015$~mag; LAFRs)  is not visible in our sample of X-ray-selected stars, except for a few outliers. We estimate that about 1\,000 of the LAFR sources identified by  \citet{distefano23} are expected to exhibit X-ray fluxes above the mean detection limit of eRASS1 if the sources are saturated in X-rays as expected given their fast rotation rates. 
        
        The fractional X-ray flux $R_X=\frac{F_X}{F_\mathrm{bol}}$ is a function of the rotation period $P$ and the color-dependent convective turnover time $\tau$. In previous studies, the fractional X-ray flux is often fitted as a broken power law of the Rossby number $R_O=\frac{P}{\tau}$ \citep[e.g.][]{pizzo03,wright11,wright18}. The convective turnover time is then estimated separately by dividing the sample in color bins; however, both estimations depend on each other. We therefore fitted both relations consistently with a two-variable function. Specifically, we described the convective turnover time as a polynomial of the color
                \begin{equation}
                        \log(\tau)(BP-RP) = \sum_{k=0}^{2} a_k(BP-RP)^k.
                        \label{equ: tau model}
                \end{equation}
        For the relation between the logarithm of the fractional X-ray flux and the Rossby number, we first adopted the classical broken power-law model 
        \begin{equation}
                \log(R_X)(P, \tau(BP-RP)) = \begin{cases}
                        \log(R_{X,\mathrm{sat}}) & \; \frac{P}{\tau(BP-RP)} < R_{O,\mathrm{sat}}\\
                        \beta\times\log\left(\frac{P}{\tau(BP-RP)}\right) & \; \frac{P}{\tau(BP-RP)} \geq R_{O,\mathrm{sat}}
                \end{cases},
        \label{equ: activity-rossby model PL}
        \end{equation}
        and alternatively, we applied a polynomial, 
        \begin{equation}
                \log(R_X)(P, \tau(BP-RP)) = \sum_{k=0}^{3} b_k\log\left(\frac{P}{\tau(BP-RP)}\right)^k.
                \label{equ: activity-rossby model poly}
        \end{equation}
        We found the best fit by varying the coefficients of Equation~\ref{equ: tau model} as well as the saturation limit $R_{X,\mathrm{sat}}$ for fast rotators, the slope $\beta$ for slow rotators, and the Rossby number $R_{O,\mathrm{sat}}$ that separates both regimes from the coefficients of Equation~\ref{equ: activity-rossby model poly},  applying the trust region reflective algorithm as implemented in \texttt{scipy.optimize.curve\_fit} \citep{scipy_curve_fit}. The approach does not allow us to determine the absolute values of the convective turnover time, and therefore we chose $a_0$ so that the turnover time of the Sun from \citet{noyes84} is recovered. Our sample contains some outliers; e.g., sources with relatively large fractional X-ray fluxes caused by X-ray variability and inaccuracies in the estimation of the rotation period, and therefore we iteratively excluded the 5\% of sources with the largest separations from the fit.
        
        \begin{table}[t]
                \caption{Best-fit parameters and standard deviations of the rotation--activity relation}
                \label{tab: rotation-activity fit}
                \centering
                \begin{tabular}{ll|ll}
                        \hline
                        \multicolumn{2}{c}{broken power law} & \multicolumn{2}{c}{polynomial} \\
                        \hline \hline
                        $a_1$ & $3.32 \pm 0.12$ & $a_1$ & $3.70 \pm 0.14$ \\
                        $a_2$ & $-0.635 \pm 0.035$ & $a_2$ & $-0.688 \pm 0.039$ \\
                        $\log(R_{O, \mathrm{sat}})$ & $-1.561 \pm 0.026$ & $b_0$ & $-4.974 \pm 0.036$ \\
                        $\log(R_{X, \mathrm{sat}})$ & $-3.0047 \pm 0.0052$ & $b_1$ & $-2.030 \pm 0.079$ \\
                        $\beta$ & $-1.110 \pm 0.031$ & $b_2$ & $-0.719 \pm 0.044$ \\
                         & & $b_3$ & $-0.0880 \pm 0.0075$ \\
                        \hline
                \end{tabular}
        \end{table}
        \begin{figure}[t]
                \includegraphics[width=\hsize]{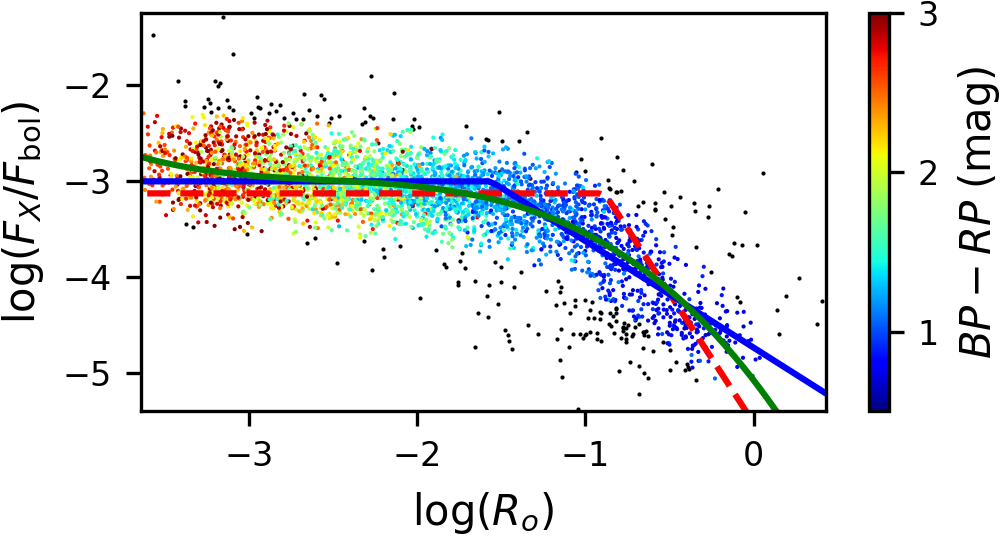}
                \caption{X-ray to bolometric flux ratio as a function of the Rossby number for the coronal eRASS1 sources with rotation period. The dashed red line shows the best-fit model of \citet{wright11} and the blue and green solid lines show the best fit of the broken power law (Equation~\ref{equ: activity-rossby model PL}) and the polynomial (Equation~\ref{equ: activity-rossby model poly}) model, respectively. The color scales with the $BP-RP$ color and the outliers excluded from the fit are shown as black dots.}
                \label{fig: rotation-activity eRASS1}
        \end{figure}
        In Table~\ref{tab: rotation-activity fit} we specify the best-fit values for Equations~\ref{equ: tau model}, \ref{equ: activity-rossby model PL}, and \ref{equ: activity-rossby model poly} together with their $1\sigma$ standard deviation errors and we compare the models of the rotation-activity relation in Fig.~\ref{fig: rotation-activity eRASS1}. To ease comparison, we also adopted the convective turnover time of the  broken power law for the polynomial model in Fig.~\ref{fig: rotation-activity eRASS1}, but not in Table~\ref{tab: rotation-activity fit}. As in previous studies \citep{pizzo03,wright11,wright18}, we find a nearly constant fractional X-ray flux for fast rotators and a decline in the X-ray activity with increasing Rossby number, but the best-fit parameters differ from those of \citet{wright11,wright18}. In our sample, the Rossby numbers decrease to later spectral types, and only very few sources bluer than $BP-RP \approx 0.9$~mag are saturated, while nearly all sources with $BP-RP \gtrsim 2.0$~mag are in the saturation regime. Hence, our estimation of the convective turnover time is less reliable in these regions. 
                
        Our estimation is representative of flux-limited samples and no attempt has been made to correct for  biases, such as those due to flares. For example, the saturation limit seems to increase to later spectral types because optically faint late-type sources are more likely to be detected if they are above the saturation limit due to intrinsic scatter and flares. A more detailed analysis of the rotation--activity relation, considering variation, upper limits, and a possible extension of the sample by pointed observations, will be the subject of a dedicated study.


        \section{Discussion of error  and caveats}
        \label{sec: Error discussion and caveats}
        The HamStar method presented here is an automated procedure optimized for the identification of the majority of coronal eRASS1 sources. For individual objects, even very prominent ones, it may provide incorrect results. The coronal probabilities are most reliable for source types with properties similar to those of the sources of our training set, because the Bayes factor is best defined in regions with many training- and control-set sources. This factor may be subject to statistical fluctuations for properties not well covered by the training and control sets. However, these regions of the parameter space contain only a few counterparts. Coronal sources whose properties overlap with the properties of the control set typically obtain lower coronal probabilities. Examples are sources detected during a large flare with X-ray fluxes exceeding the saturation limit. In general, the identifications of bright X-ray sources with a high detection likelihood are more reliable than low likelihood detections.
        
        The goal of the present work is to identify the coronal content of eRASS1, and we do not provide associations for most of the eRASS1 sources, because the majority of them are expected to be associated with an AGN (cf. Salvato et al., in prep.), and these sources are excluded from our sample of eligible coronal counterparts. The fact that we can only  find the correct identification for a number of eRASS1 sources is considered in our prior by the coronal fraction $p_c$ as a sample property. However, the information about alternative, possibly prominent noncoronal counterparts for individual sources is not taken into account for the estimation of the coronal probability. We flag sources with an alternative noncoronal counterpart in NWAY to make users aware of such cases. Furthermore, our identifications are limited by the completeness of \textit{Gaia} DR3 and coronal eRASS1 sources not part of \textit{Gaia} DR3 or with a missing or uncertain parallax cannot be identified by HamStar unless they have a \textit{Hipparcos} parallax. Spurious parallax or photometric measurements in \textit{Gaia} DR3 might also influence our identifications and derived properties. As our counterparts are rather bright in the \textit{Gaia} context, the number of missing or spurious \textit{Gaia} sources is expected to be very small. We do not consider extinction and reddening effects on our counterparts because this information is not available to all sources and the influence is generally small for the nearby coronal eRASS1 sources. Despite our filter criteria, the eligible coronal counterparts contain some noncoronal sources, which are mainly CVs. We flagged obvious cases, but the X-ray emission of a small number of our identifications might be produced by accretion processes and not in a corona.
        
        The X-ray flux of the coronal eRASS1 sources has a mean error of about 35\%. The error on the X-ray flux therefore provides the largest contribution to the error in the identifications and the derived properties, such as the X-ray luminosity. In comparison, the application of the parallax inversion causes a deviation toward more complex distance estimations of less than 5\% for most sources.

        \section{Conclusions and outlook} 
        \label{sec: Conclusions}
        We present the coronal content of the eRASS1 catalog obtained with our HamStar method, which is designed to identify coronal X-ray sources. HamStar considers geometrical properties, such as angular separation and positional accuracy, as well as physical properties of the underlying sources, such as the fractional X-ray flux, color, and the distances of eligible coronal \textit{Gaia} DR3 counterparts. We obtain identifications for 139\,700 eRASS1 sources but the properties of a few objects suggest that their X-ray emission is produced in an accretion process and not in a corona. The remaining 138\,800 eRASS1 sources with a coronal counterpart in the western hemisphere is presently the largest sample of coronal X-ray emitters available.  Overall, we identify 15.5~\% of the point-like eRASS1 main sources to be coronal, but the coronal fraction strongly varies between the Galactic poles and the Galactic plane, where the coronal fraction reaches almost 50\%. For sources flagged in the eRASS1 catalog, we found fewer coronal counterparts, which confirms that many of these sources are spurious detections. 
        
        For every eRASS1 source, we provide a probability  of it being coronal, which allows us to select samples of different purity. At a probability cutoff that recovers the expected number of coronal sources, a completeness and reliability of about 91.5~\% is reached. The quality of the identifications is confirmed by sources detected with \textit{Chandra} at a higher positional uncertainty. We also compared our identifications with alternative counterparts provided by NWAY (Salvato et al. in prep.), which is designed to primarily identify extragalactic sources. The results from both identification schemes agree within the expectations. While HamStar provides better results for the coronal identifications, NWAY supplies alternative extragalactic counterparts that are not considered by HamStar.
        
        The properties of the coronal eRASS1 sources are generally similar to those of the sample of coronal sources from RASS, but because of the larger sample size, we find more objects from the rarer populations. The secondary counterpart for more than 10\% of the coronal eRASS1 sources has a similar parallax to the primary counterpart, and therefore the two probably belong to a wide binary system or the same stellar association. Many of the counterparts are also known close binaries that are unresolved in \textit{Gaia} DR3 and the number of binaries with periods of less than 10 days is enhanced in our X-ray-selected sample. The potential of the coronal eRASS1 identification is also demonstrated by a cross-match with rotation periods from \citet{distefano23}; this results in 3\,721 sources in the largest available sample of X-ray-detected sources with a known rotation period. We estimated convection turnover times and fitted the rotation--activity relation for our flux-limited sample with a broken power law and a polynomial model. 
        Contrary to expectations, we do not find the population of fast rotators with low amplitudes described by \citet{distefano23}, which might indicate that the rotation periods from \textit{Gaia} DR3 are either wrong for these sources or that there is a population of fast rotators that are not saturated in X-rays. An even larger sample of X-ray-emitting rotators will be obtained by analyzing rotational modulations with TESS (Schmitt et al. in prep.). 
        
        Future \textit{eROSITA} all-sky surveys will provide deeper exposures and more coronal X-ray emitters; these will allow detailed studies of the short- and long-timescale  X-ray variability of stars; that is, from minutes to years. This will further improve our understanding of coronal X-ray emission.

        \begin{acknowledgements} 
                We thank Andrea Merloni for very helpful comments and suggestions.
                
                SF gratefully acknowledge supports through the Integrationsamt Hildesheim and the ZAV of Bundesagentur f\"ur Arbeit, SC by DFG under grant CZ 222/5-1, JR by DLR under grant 50 QR 2105, and PCS by DLR under grant 50 OR 1901 and 50 OR 2102. SF thanks Gabriele Uth and Maria Theresa Lehmann for their support.

                This work is based on data from eROSITA, the soft X-ray instrument aboard SRG, a joint Russian-German science mission supported by the Russian Space Agency (Roskosmos), in the interests of the Russian Academy of Sciences represented by its Space Research Institute (IKI), and the Deutsches Zentrum für Luft- und Raumfahrt (DLR). The SRG spacecraft was built by Lavochkin Association (NPOL) and its subcontractors, and is operated by NPOL with support from the Max Planck Institute for Extraterrestrial Physics (MPE).
                
                The development and construction of the eROSITA X-ray instrument was led by MPE, with contributions from the Dr. Karl Remeis Observatory Bamberg \& ECAP (FAU Erlangen-Nuernberg), the University of Hamburg Observatory, the Leibniz Institute for Astrophysics Potsdam (AIP), and the Institute for Astronomy and Astrophysics of the University of Tübingen, with the support of DLR and the Max Planck Society. The Argelander Institute for Astronomy of the University of Bonn and the Ludwig Maximilians Universität Munich also participated in the science preparation for eROSITA.
                
                The eROSITA data shown here were processed using the eSASS/NRTA software system developed by the German eROSITA consortium.

                This work has made use of data from the European Space Agency (ESA) mission
                {\it Gaia} (\url{https://www.cosmos.esa.int/gaia}), processed by the {\it Gaia}
                Data Processing and Analysis Consortium (DPAC,
                \url{https://www.cosmos.esa.int/web/gaia/dpac/consortium}). Funding for the DPAC
                has been provided by national institutions, in particular the institutions
                participating in the {\it Gaia} Multilateral Agreement.

                This research has made use of data obtained from the Chandra Source Catalog, provided by the Chandra X-ray Center (CXC) as part of the Chandra Data Archive.

                This research has made use of the SIMBAD database, operated at CDS, Strasbourg, France.

                This research has made use of the VizieR catalogue access tool, CDS,
                Strasbourg, France (DOI : 10.26093/cds/vizier). The original description 
                of the VizieR service was published in 2000, A\&AS 143, 23
        \end{acknowledgements}

        \bibliographystyle{aa} 
        \bibliography{mybib}

        \begin{appendix}
                \section{Bayes maps}
                \label{sec: Bayes maps}
\begin{figure*}[t]
                        {\includegraphics[width=8.5cm]{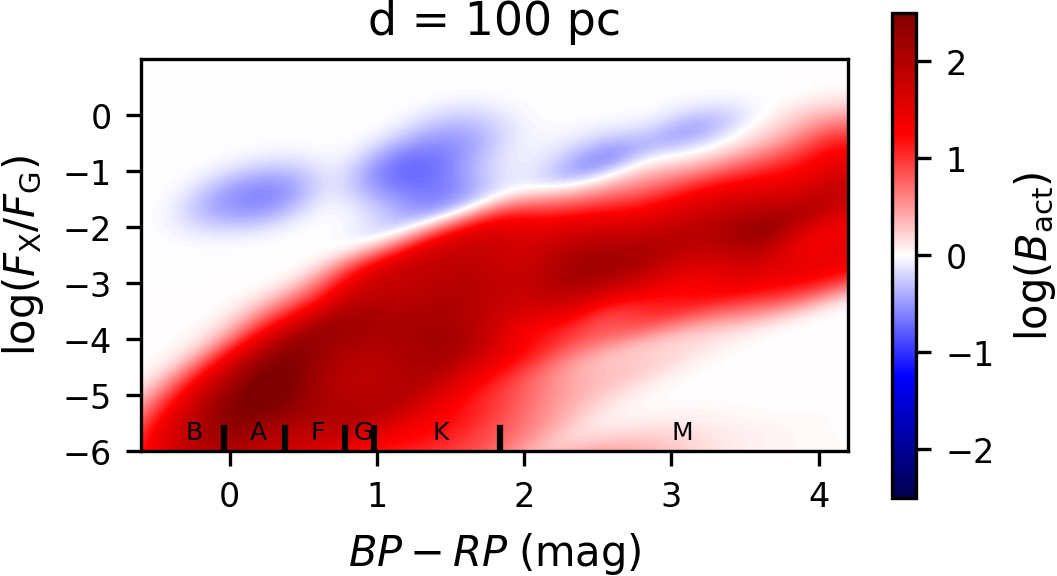}}
                        \hfill
                        {\includegraphics[width=8.5cm]{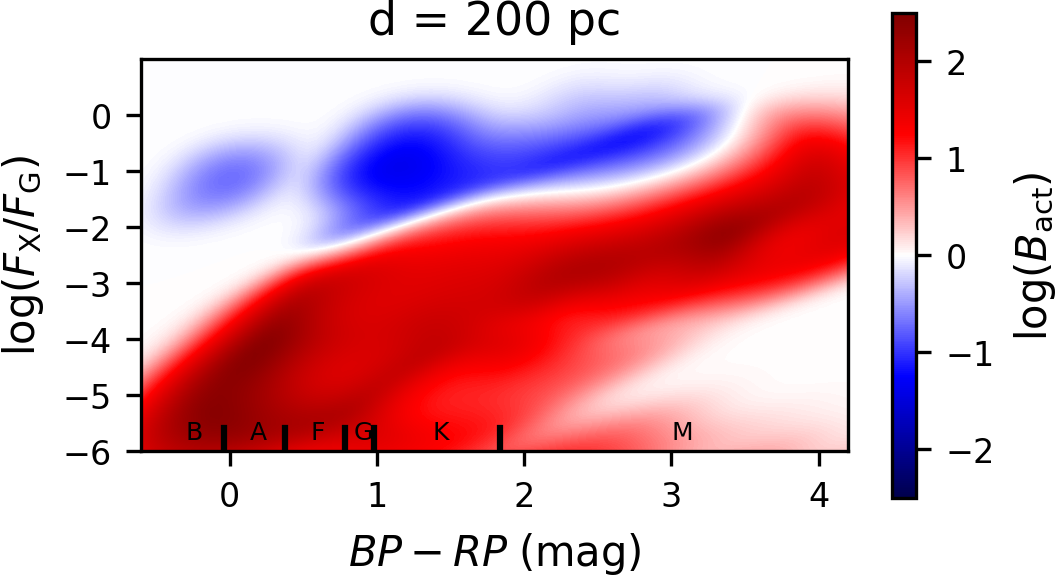}}
                        \par\bigskip
                        {\includegraphics[width=8.5cm]{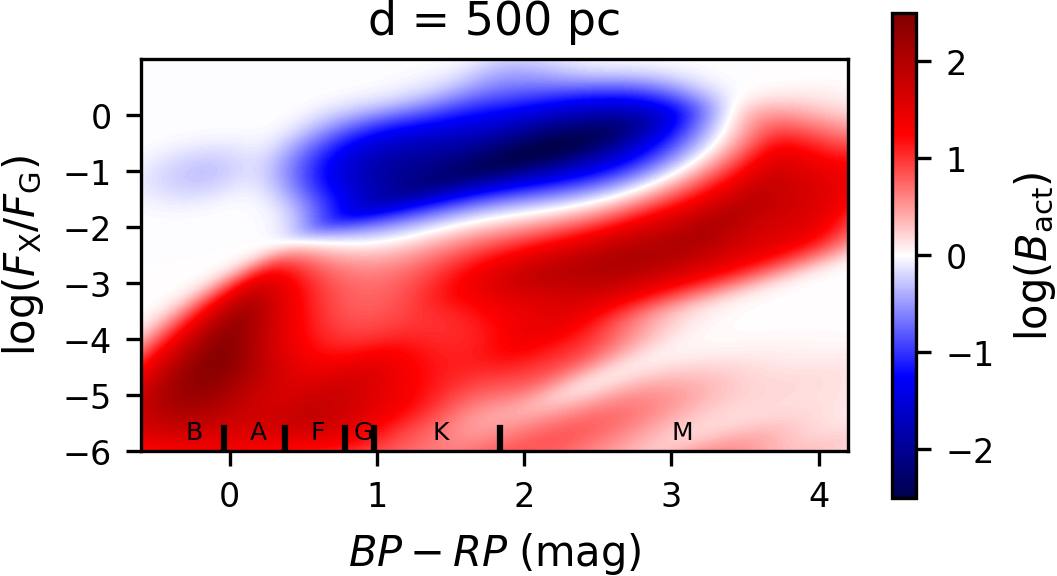}}
                        \hfill
                        {\includegraphics[width=8.5cm]{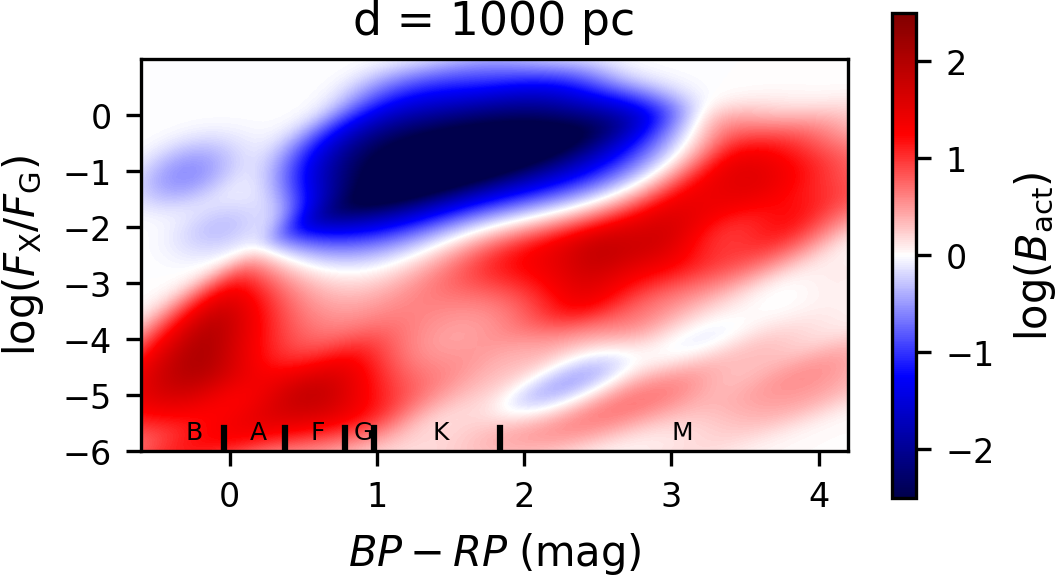}}
                        \caption{Bayes maps for sources at different distances. The color scales with the strengths of the weighting. } 
                        \label{fig: Bayes maps}
                \end{figure*}

                To illustrate how the properties of the counterparts are weighted, we show in Fig.~\ref{fig: Bayes maps} the Bayes maps of the X-ray over G-band flux ratio as a function of the $BP-RP$ color for counterparts at different distances. Counterparts with small fractional X-ray fluxes generally obtain large Bayes factors leading to an increase in the geometrically estimated matching probability, while the Bayes factor for sources above the saturation limit is smaller than unity and the matching probability is downweighted. Although the shape of the Bayes map is similar at all distances, the area where the probability is downweighted increases at the cost of the region with upweighted probability at larger distances, and similarly, the strength of the downweighting increases and decreases for the upweighting. Some regions of the Bayes map seem to be unphysical; for example, the small downweighted area for early M-type sources with $\log(F_X/F_G) \approx -5$ or the increase in the Bayes factors at very large fractional X-ray fluxes. This is caused by low statistics, because some regions are not populated by training or control set sources. Due to the very small number of counterparts with these properties, this has very little to no effect on our identifications. 
                
                The Bayes maps shown in Fig.~\ref{fig: Bayes maps} apply to sources with Galactic coordinates $l>300^\circ$ and $10^\circ < |b| < 20^\circ$. At other coordinates, the Bayes maps are generally similar but the dependence on the distance is stronger near the Galactic center than at the poles. 

                \section{Column description}
                \label{sec: column description}
                In Table~\ref{tab: column description} we describe the columns of the published catalog. 
                \begin{table*}
                        \caption{Description of the columns in the published catalog}
                        \label{tab: column description}
                        \centering
                        \begin{tabular}{p{2.5cm} p{3.5cm} p{10cm}}
                                \hline
                                Column & Unit & Description \\
                                \hline \hline
                                \texttt{ERO\_IAUNAME} & & IAU name of the eRASS1 source (\texttt{IAUNAME}) \\
                                \texttt{ERO\_MJD} & modified julian date & Epoch of the observation of the eRASS1 source (\texttt{MJD}) \\
                                \texttt{ERO\_POS\_ERR} & arcsec & Positional uncertainty of the eRASS1 source (\texttt{POS\_ERR}) \\
                                \texttt{CTP\_RANK} & & Rank of the counterpart that increase for less likely alternative counterparts \\
                                \texttt{CTP\_ID} & & Identifier of the counterpart (\texttt{source\_id} if \textit{Gaia} source, otherwise combination of \texttt{TYC1, TYC2, TYC3}) \\
                                \texttt{CTP\_SEP} & arcsec & Proper motion corrected angular separation between eRASS1 source and counterpart \\
                                \texttt{p\_coronal} & & Probability of the eRASS1 source to be coronal \\
                                \texttt{p\_ij} & & Probability of the counterpart to be associated to the eRASS1 source \\
                                \texttt{CTP\_RA} & deg & Proper motion corrected right ascension of the counterpart \\
                                \texttt{CTP\_DEC} & deg & Proper motion corrected declination of the counterpart \\
                                \texttt{Fx} & erg cm$^{-2}$ s$^{-1}$ & X-ray flux in the 0.2 -- 2.3~keV band assuming a coronal spectrum \\
                                \texttt{G} & mag & G band magnitude of the counterpart \\
                                \texttt{BP\_RP} & mag & $BP-RP$ color of the counterpart \\
                                \texttt{PLX} & mas & Parallax of the counterpart \\
                                \texttt{SIMBAD\_NAME} & & Name of the Simbad identification if available \\
                                \texttt{SIMBAD\_OTYPE} & & Classification in the Simbad database if available \\
                                \texttt{TRAIN\_SET} & & True if counterpart to an eRASS:4 training-set source \\
                                \texttt{CORONAL} & & False if the properties are untypical for coronal X-ray emitters \\
                                \texttt{OB\_STAR} & & True for likely reddened OB stars \\
                                \texttt{FLAG\_OPT} & & True if source is affected by optical loading, adopted from the eRASS1 catalog \\
                                \texttt{NWAY\_DIFFCTP} & & True if NWAY provides a different best counterpart \\
                                \hline
                        \end{tabular}
                \end{table*}
        \end{appendix}

\end{document}